\documentclass[aps,prb,twocolumn,float,amsmath]{revtex4}
\usepackage{xcolor}
\usepackage{graphicx}
\usepackage{amsmath,bm,epsfig}
\definecolor{g-blue}{rgb}{0.83,0.95,1}
\usepackage{bm}
\usepackage{amssymb}
\usepackage{amsfonts}
\usepackage{amsmath}
\usepackage{graphicx} 
\usepackage{booktabs}
\usepackage{array}
\usepackage{threeparttable}
\usepackage{multirow}
\usepackage{soul}

\let \= \equiv  \let\*\cdot \let\~\widetilde \let\^\widehat \let\-\overline

 \def\1{\bm1} 

\def\<{\left\langle}    \def\>{\right\rangle}
\def\({\left(}          \def\){\right)}
\def \[ {\left [} \def \] {\right ]}
\newcommand{\Eq}[1]{Eq.\,(\ref{#1})}%%  requires \eq{label}
\newcommand{\Eqs}[1]{Eqs.\,(\ref{#1})}%%  requires \eq{label}
\newcommand{\Table}[1]{Table.\,\ref{#1}}%%  requires \Fef{label}
\newcommand{\Fig}[1]{Fig.\,\ref{#1}}%%  requires \Fef{label}
\newcommand{\Figs}[1]{Figs.\,\ref{#1}}%%  requires \Fef{label}
\newcommand{\Sec}[1]{Sec.\,\ref{#1}}%%  requires \Fef{label}
\newcommand{\Refn}[1]{Ref.\,\cite{#1}}%%  requires \Fef{label}
\newcommand{\Refs}[1]{Refs.\,\cite{#1}}%%  requires \Fef{label}
\def\Fbox#1{\vskip1ex\hbox to 8.5cm{\hfil\fboxsep0.3cm\fbox{%
			\parbox{8.0cm}{#1}}\hfil}\vskip1ex\noindent}  %%  {TEXT}

\renewcommand{\sb}[1]{_{\text {#1}}}  %% sub-   for lower case
\renewcommand{\sp}[1]{^{\text {#1}}}  %% super- for lower case
\def\Sb#1{_{\scriptscriptstyle\rm{#1}}}
\def\He4 {$^4$He~}
\newcommand{\B}[1]{{\bm{#1}}}%% Bold Roman & Greek Lower & Upper Case
\newcommand{\C}[1]{{\mathcal{#1}}}    %%   Calligrapfic Upper case

\def\red#1{\textcolor{red}{#1}}

\begin{document}

\title{Eulerian and Lagrangian second-order statistics of superfluid ${^4}$He grid turbulence}

\author{Y. Tang, W. Guo}
\affiliation{National High Magnetic Field Laboratory, 1800 East Paul Dirac Drive, Tallahassee, FL 32310, USA}
\affiliation{Mechanical Engineering Department, Florida State University, Tallahassee, FL 32310, USA}

\author{V.S. L'vov, A. Pomyalov}
\affiliation{Department of Chemical and Biological Physics, Weizmann Institute of Science, Rehovot 76100, Israel}

\begin{abstract}
 We use particle tracking velocimetry to study Eulerian and Lagrangian  second-order statistics of superfluid $^4$He grid turbulence. The Eulerian energy spectra  at scales larger than the mean distance  between quantum vortex lines behave classically with close to Kolmogorov-1941 scaling and are almost isotropic.  The Lagrangian second-order structure functions and frequency power  spectra, measured at scales comparable with the intervortex distance, demonstrate a sharp transition from nearly-classical behavior to  a regime dominated by the  motion of quantum vortex lines. Employing the homogeneity  of the flow, we verify a set of relations that connect various second-order statistical objects that stress different aspects of turbulent behavior, allowing a multifaceted analysis. 
 We use the two-way bridge relations between  Eulerian energy spectra and second-order structure functions
 to reconstruct the energy spectrum from the known second-order velocity structure function and vice versa. The Lagrangian frequency spectrum reconstructed from the measured Eulerian spectrum using the Eulerian-Lagrangian bridge differs from the measured Lagrangian spectrum in the quasi-classical range which calls for further investigation.    
\end{abstract}
\maketitle
%\tableofcontents

 \section*{ Introduction}
The statistical description of turbulent  flows follows two distinct approaches. In the Eulerian approach, the main objects of interest are velocity differences for various spatial separations. In the Lagrangian approach, the fluid particles are followed along their trajectories, and the main focus is the velocity differences at sequential time moments.

These two approaches complement each other. The Eulerian approach to turbulence is more convenient for phenomena dominated by large scale motions, such as wall-bounded turbulence. The Lagrangian description of turbulence has unique physical advantages that are especially important in studies
of phenomena dominated by small-scale motions or many-point correlation functions, like turbulent mixing and particle dispersion.

The Eulerian second-order statistics, namely the velocity structure functions $S(r)$ (square of the spatial velocity increments across a separation $r$) and the energy spectra of turbulent velocity fluctuations in the wavenumber $k$-space $E(k)$, have been studied in depth over years\cite{1953-Bat,1971-CBB,1963-Cor,1935-Tay,1975-Ten, Frisch,ToschiBoden09,1974-SC}. The typical experimental studies use one-probe measurements (e.g. by hot-wire anemometer), which provide researchers with the time dependence of the Eulerian velocity $\B u(\B r\sb{pr},t)$ at the position of the probe $\B r\sb {pr}$. In the presence of large mean velocity (such as wind in atmospheric measurements), Taylor hypothesis\cite{Taylor-1938} of frozen turbulence allows one to transform these data into coordinate dependence $\B u(x,t)$ in the streamwise direction $ \widehat{\B  x}$ practically at the same moment of time $t$.

One-probe measurements cannot give information about a velocity $\B v(\B r_0|t)$ of a Lagrangian tracer (a mass-less particle, swept by turbulent velocity field without slippage) positioned at  $\B r=\B r_0$ for $t=t_0$.
% To the best of our knowledge for the first time
The experimental study of Lagrangian velocity $\B v(\B r_0|t)$ was done by Snyder and Lumley who provided\cite{Snyder71} the first systematic set of particle-tracking velocity measurements from the optical tracking of tracer particles in wind-tunnel grid turbulence. The technique is known as particle tracking velocimetry (PTV), when velocities are concerned, or Lagrangian particle tracking, when the position and the acceleration are also determined. Particles can be passive tracers that approximate the Lagrangian motion of fluid elements, have inertia, or have a size larger than the smallest scales of the flow.

Extensive experimental and Direct Numerical Simulation (DNS) studies on the spatial and temporal velocity correlations [see, e.g. \Refs{Yeung02,Yeung06,1974-SC,Oullette,Berg}]  have been conducted in turbulent flows in classical fluids. There is a strong desire in increasing the interval of scales where the behavior of the fluid structure functions and spectra is universal, which can be achieved by using a fluid material with low kinematic viscosity. Liquid helium is known for its extremely small viscosity. However, unlike the classical fluids, only a few experimental techniques are available for liquid helium and most of them do not allow extracting the local information on the turbulent fluctuations. Recently, flow visualization\cite{Marakov15,Ladik,Gao17,WG1,WG-2014,Bewley-2006,Bao-2018}  was extended to liquid helium, which has provided invaluable information about its rich fluid dynamical behaviors. 
	
In particular, there has been an increasing interest in studying turbulent flows in superfluid phase\cite{1,2,3,4}  of liquid helium (i.e., He II), which occurs below a critical temperature $T_{\lambda}=2.17$ K.  At these conditions,  fluid vorticity may be divided into two parts.  A quantized part forms a quantum  ground state, while  thermal excitations represent a normal fluid with continuous vorticity. The vorticity quantization results\cite{5} in the creation of thin quantum vortex lines of fixed circulation that interact with the normal fluid via mutual friction force, forming dense tangles.  

Large-scale hydrodynamics of such a system is usually described by a two-fluid model, interpreting $^4$He as a mixture of two coupled fluid components: an inviscid superfluid and a viscous normal fluid. The tangle of quantum vortexes mediates the interaction between fluid components via mutual friction force\cite{1}.

There are many indications\cite{3,6,7,8,Maurer-QC,Stalp-QC,Emil-2018} that the mechanically driven turbulence in the superfluid $^4$He is similar to the turbulence in the classical flows. In this paper, we  apply the particle tracking velocimetry to study the second order statistics in the grid superfluid $^4$He turbulence.
We confirm the near-Kolmogorov\cite{Frisch} scaling of the Eulerian energy spectra, which are also almost isotropic. The small-scale statistics studied using Lagrangian structure functions and energy spectra deviate from the classical behavior, illustrating a clear transition from a random fluid motion to a motion dominated by the velocity of individual vortex lines at scales comparable with the intervortex spacing.

The paper is organized as follows. In \Sec{s:bgr} we summarize some important information about the second-order statistics and the relation between the Eulerian and Lagrangian energy spectra in the classical turbulence. In \Sec{s:2} we describe the experimental techniques (\Sec{ss:exp}) and  the method to extract the 2D velocities (\Sec{ss:Eul}) and the Eulerian energy spectra from the PTV data (\Sec{ss:method}). \Sec{s:spectra} is devoted to the experimental results and their discussion. We start with the Eulerian statistics in \Sec{ss:Euler}. We discuss various forms of the energy spectra as well as relations between the spectra and the structure and correlation functions. Then we turn to the Lagrangian statistics (\Sec{ss:Lagr}) where we consider the second-order structure functions (\Sec{ss:LagrS}) and the energy spectra (\Sec{ss:LagrE}). In Conclusions we summarize our findings.

\section{\label{s:bgr}Analytical background}
The goal of this Section is to recall well-known definitions of the second-order statistical objects in the Eulerian and Lagrangian description of turbulence, to introduce their notations, and to remind relations between them.
\subsection{Second-order statistical description of turbulence}
The complete statistical description of spatially and temporally homogeneous random processes with Gaussian statistics can be done on the level of their quadratic (second-order) statistical objects. Statistics of developed hydrodynamic turbulence is not Gaussian. Nevertheless, the most basic characteristics of turbulence, the energy distributions among spatial and temporal scales, are given by their second-order objects, i.e., the energy spectra and the structure functions.

\subsubsection{\label{sss:bridges} Bridge equations for Eulerian structure functions  and energy spectra}
The  Eulerian approach to the statistics of homogeneous stationary turbulence is applied to the  field of the turbulent velocity fluctuations $\B u(\B r, t)$ with zero mean $\langle\B u(\B r, t)\rangle=0$. Here $\langle\dots \rangle$ is a ``proper" averaging, which can be understood as the time averaging in stationary turbulence, the ensemble averaging over realizations in experiments or numerical simulations, \textit{etc}.

As the basic objects in the second-order statistical description of turbulence in the Eulerian framework, we can take the simultaneous two-point correlation function (covariance)
\begin{subequations}\label{1}
\begin{equation}\label{1a}
C_{\alpha\beta}(\B r )\=\< u_\alpha ( \B r+\B r',t) u_\beta (\B r' ,t)\>
\end{equation}  
and its Fourier transform
\begin{equation}\label{1b}
F_{\alpha\beta}(\B k)\= \int d^3  r \,C_{\alpha\beta}(\B r) \exp (i \B k \cdot \B r)\ .
\end{equation} 
Here the indices $\alpha, \ \beta= \{x, \ y,  z \}$ denote Cartesian coordinates.  
The inverse transform to \Eq{1b} takes the form:
\begin{equation}\label{1c}
C_{\alpha\beta}(\B r) \= \int   \,F_{\alpha\beta}(\B k) \exp (-i \B k \cdot \B r)\frac{d^3 k}{(2\pi)^3}\ .
\end{equation} 
\end{subequations}
Alternatively, the function $F_{\alpha\beta}(\B k)$ can be expressed in terms of the Fourier transform of the velocity field
 \begin{subequations}\label{2}
 	\begin{equation}\label{2a}
 	 \B u(\B k)= \int d^3  r \,\B u(\B r) \exp (i \B k \cdot \B r)\ .
 	\end{equation}  
 as follows
 	\begin{align}\begin{split}\label{2b}
 	 (2\pi)^3 \delta(\B k-\B k')F_{\alpha\beta}(\B k)&= \< u_\alpha (\B k) u_\beta^* (\B k')\>\,, \\
 	  	  \sum_\alpha F_{\alpha\alpha}(\B k)&\=F_{\alpha\alpha}(\B k)\=F (\B k)\ .
 \end{split}	\end{align} 
 Hereafter we adopt Einstein summation over repeated indices and skip  for  shortness the repeated vector indices in the resulting  traces of all tensors, e.g.  $F_{\alpha\alpha}(\B k)\=F (\B k)$, $C_{\alpha\alpha}(\B r)\= C(\B r)$, etc.

 One sees that $  F (\B k)$ in \Eq{2b} describes the turbulent energy spectrum -- the density of twice the turbulent kinetic energy per unit mass in the wave-vector three-dimensional space $\B k$ . In particular, it follows from \Eqs{1} that
 \begin{eqnarray}\label{2c}
 \int F (\B k) \frac{d^3   k}{(2\pi)^3}= C (0)= \< |\B u(\B r)|^2\>  \ .%, \  C (\B r)\= C_{\alpha\alpha}  (\B r)  .~~
 \end{eqnarray}
\end{subequations}
In their turn, the correlations\,\eqref{1a} are simply connected to the second-order structure functions of the velocity differences $ S(\B r) $ in the homogeneous turbulence 
 	\begin{align} \begin{split}  \label{3}
S(\B r)  & \=    \langle  | \B u  (\B r+\B r')-\B u  (  \B r')|^2\rangle = 2\,\big[ C(0)-C(\B r)\big ]\ .\\
 \end{split}	\end{align}

   The equation\,\eqref{1}, the one-dimensional (1D) version of which is known in the theory of stochastic processes as Wiener-Khinchin-Kolmogorov theorem\cite{Khin}, can be considered in the theory of turbulence as two-way  bridge equations which allows one to find the  Eulerian correlation   function $C(\B r)$ if one knows the energy spectrum $F(\B k)$ and vice versa: to  find   $C(\B r)$ from known  $F(\B k)$.  
 In the isotropic turbulence, where $F(\B k)$ and $C(\B r)$ depend only on the wavenumber $k=|\B k|$ and $r=|\B r|$, these equations may be simplified:  
  \begin{subequations}\label{4a} 	\begin{eqnarray}
	\label{4Aa}
C (  r) & =& \int \limits_0^\infty  \,F ( k) \frac {\sin (  k    r)}r \frac{k d k}{ 2\pi ^2}\,,   \\    \label{4Bb}
  F ( k) &=&  4 \pi \int\limits_0^\infty    \,C ( r) \frac {\sin (  k    r)}k \  r dr  \ .
  \end{eqnarray}\end{subequations}
For 1D case, one should replace $\int d^3 r$ in \Eq{1b} by $\int_{-\infty}^\infty  d r $ and replace $\int d^3 k$ in \Eq{1c} by $\int_{-\infty}^\infty  d k$. This gives  the 1D $  k\!  \Leftrightarrow\!    r $-bridge equations
   \begin{subequations}\label{4b}  	\begin{eqnarray} 
\label{4A}
C (  r) & = &\frac 1\pi  \int \limits_0^\infty  \,E ( k) \cos(k\, r) d k  \,,   \\  \label{4B}
C ( 0) & = &\frac 1\pi  \int \limits_0^\infty  \,E ( k)  d k  \,,   \\     \label{4C}
S (  r) & = &\frac 4\pi  \int \limits_0^\infty  \,E ( k) \sin^2\Big(\frac {k\, r}2\Big) d k  \,,   \\  \label{4D}
E ( k) &= & 2   \int\limits_0^\infty    \,C ( r) \cos(k\, r) dr  \ .
 \end{eqnarray} 
\end{subequations}
In the rest of the paper  $E ( k)$ denotes  1D Eulerian energy spectrum. We should stress here that the assumption of space homogeneity and stationarity  of the turbulent flow is essential in derivations of \Eqs{1},  \eqref{4a} and \eqref{4b}.

\subsubsection{Bridge equations for Lagrangian structure functions and frequency power spectra}

 Similar to the Eulerian approach to statistics of turbulence, important   information in the Lagrangian framework is contained in the 2$\sp{nd}$-order statistical objects: the Lagrangian correlation $\C C(\tau)$ and structure functions $\C S(\tau)$: 
	\begin{subequations}\label{5}
		\begin{align}\begin{split}\label{5a}
		\C C (\tau)&\= \langle \B v(\B r_0|t+\tau)\cdot   \B v(\B r_0|t) \rangle\,,\\	
		\C S (\tau)&\= \langle |\B v(\B r_0|t+\tau)-  \B v(\B r_0|t)|^2\rangle\,,
	\end{split}	\end{align}
	together with  the Lagrangian   energy spectrum $\C E(\omega)$, defined via $\B v(\B r_0|\omega)$, the  Fourier transform of   $\B v(\B r_0|t)$:
	\begin{align}\begin{split}\label{5b}
	(2\pi)  \delta(\omega -\omega')\C E(\omega)&\= \< \B v(\B r_0|\omega)\cdot   \B v^*(\B r_0|\omega')\>\,, \\
	  \B v(\B r_0|\omega)&\= \int    \,\B v(\B r_0|t) \exp (i \omega t)\, d t \ .
	\end{split}	\end{align} \end{subequations}

Similar to \Eqs{1b} and \eqref{4b}, $\C E(\omega)$ is the Fourier transform of $\C C(\tau)$: 
		\begin{subequations}\label{6}
	\begin{equation}\label{7a}
	 \C E (\omega)  =  2   \int\limits_0^\infty    \,\C C (\tau) \cos(\omega \tau ) dt\ .
	 	\end{equation}
	 	This relation can be considered as the $\tau \Rightarrow \!\omega\!$-bridge equation,  which expresses $\C E (\omega)$ in terms of $\C C(\tau)$ and $\C S(\tau)$
	 	\begin{equation}\label{7b}
	 	\C S(\tau)= 2 \,\big[\C C(0)-\C C(\tau)\big]\ . 
	 	\end{equation}
	 	The inverse $\omega\Rightarrow \!\tau\! $-bridge has the form similar to \Eq{4b}:
	 		\begin{equation}\label{7c}
		\C S(\tau)= \frac2\pi \int \sin^2\Big (\frac{\omega \, \tau}2\Big ) \C E(\omega) d \omega\ .
		\end{equation}
	\end{subequations}

\begin{figure*}[htp]\begin{tabular}{ccc}
		(a)&(b)&(c)\\
		\includegraphics[scale=0.4]{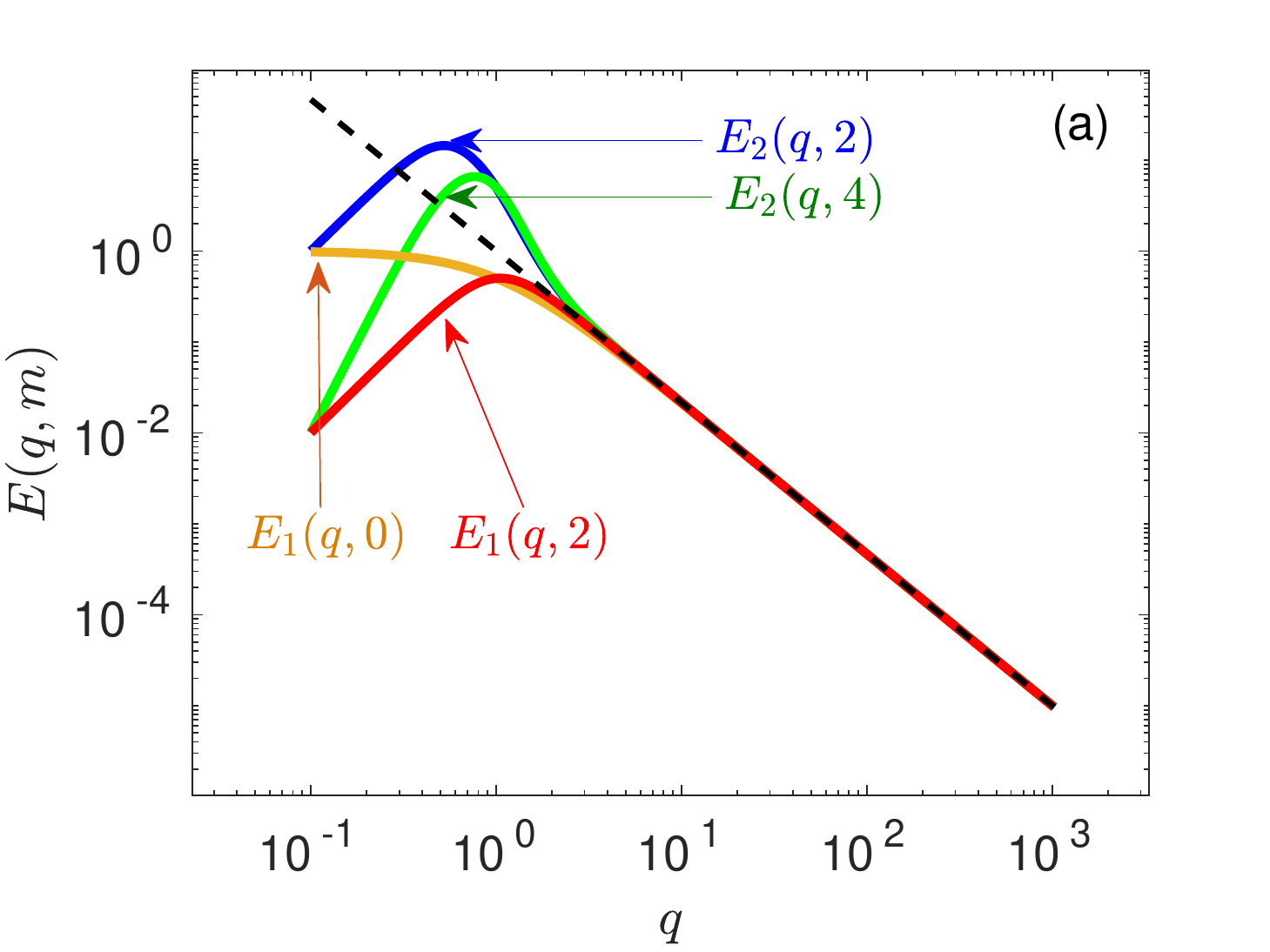} &
		\includegraphics[scale=0.4]{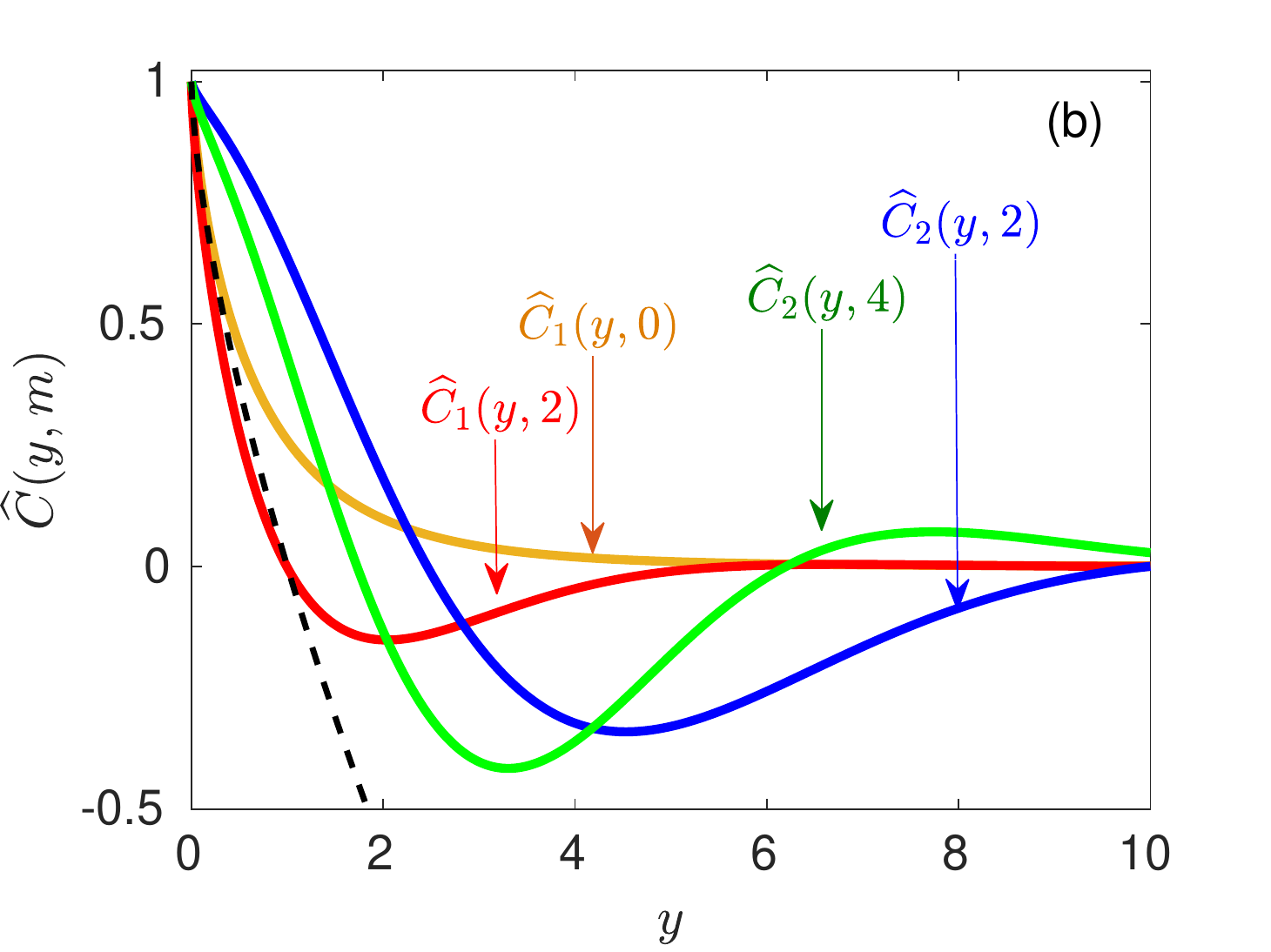}&
		\includegraphics[scale=0.4]{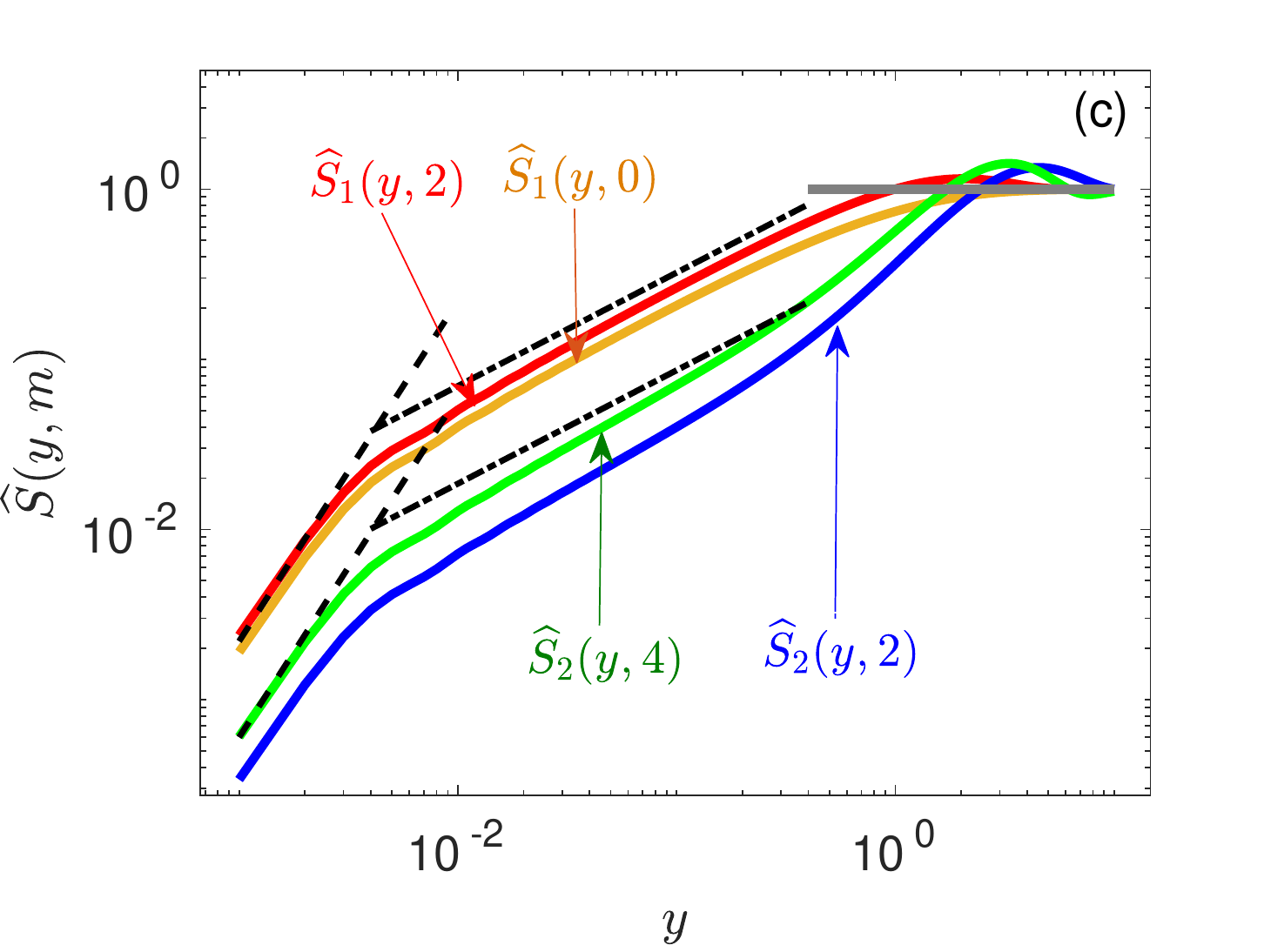}\\
	\end{tabular}

	\caption{\label{f:1}  (a)  Energy spectra  \Eq{mod}: $E_1(q,m)$  for $m=0$ (orange line) and $m=2$ (red line), and  $E_2(q,m)$  for $m=2$  (blue line) and $m=4$ (green line).  The black dashed line corresponds to  K41 scaling $q^{-5/3}$. (b)   Normalized correlation functions $\widehat{C}(y,m)$, found from the corresponding spectra via bridges \Eqs{4b}.  The color code is same as in (a).   The black dashed line denotes the  asymptote $1-y^{2/3}$. Note the linear scale. (c) The normalized structure functions  $\widehat S_n(y)=1-\widehat{C}(y,m)$. The dashed  lines denote the viscous asymptote  $\widehat S(y,m)\propto y^2$, the dot-dashed lines denote K41 scaling $\widehat S(y,m)\propto y^{2/3}$, the  horizontal thick gray line  marks the saturation level   $\widehat S(y,m) =1$.  Note the logarithmic scale.
	}	
\end{figure*}

\subsubsection{Kolmogorov-1941   scaling }

 The shapes of the   energy spectra $E(k)$ and $\C E(\omega)$ together with the  shapes of the  structure functions  $S(r)$ and $\C S(\tau)$  in the  classical  isotropic fully developed turbulence were the subject of intensive studies over  decades,
	see e.g. \Refs{1953-Bat,1971-CBB,1963-Cor,1935-Tay,1975-Ten, Frisch,ToschiBoden09,1974-SC}.   Up to relatively small intermittency corrections, these objects obey the  Kolmogorov-1941 (K41) scaling\,\cite{Frisch}:
	\begin{subequations}
		\begin{eqnarray}\label{K41}
		E\Sb{K41}(k)  \simeq  \frac{\varepsilon^{2/3}}{k^{5/3}}\,, &\quad& \C E\Sb{K41}(\omega)\simeq \frac{\varepsilon }{\omega^{2}}\,, \\
		S\Sb{K41}(r)  \simeq   (\varepsilon\,r)^{2/3}\,, &\quad&  \C S\Sb{K41}(\tau)\simeq  \varepsilon \, \tau\ .
		\end{eqnarray}
	\end{subequations}
	Here  $\varepsilon$ is the rate of dissipation of kinetic energy per
	unit mass.

\subsubsection{\label{s:bridge}  Eulerian-Lagrangian bridge equation for the  energy spectra  }

 The first attempt to relate Eulerian and Lagrangian statistics was made by Corrsin\cite{4}. He connected the Eulerian and Lagrangian correlation functions $C(r) $ and $\C C(\tau)$ via mean single-particle  Green's function. However, the experimental data\cite{5,5a} show that the $C(r) $ and $\C C(\tau)$  behave fundamentally differently, thereby seriously questioning the rationale of the conjecture. Recently,  Kamps,  Friedrich,  and  Grauer\cite{6} presented a formal connection between Lagrangian and Eulerian velocity increment distributions involving so-called transition probabilities, which so far were calculated only for two-dimensional turbulence. Unfortunately, this connection cannot be used in the practical analysis of three-dimensional turbulence. For this purpose we will use relation
\begin{equation}\label{bridge}
\C E(\omega)= 2\pi\int \exp\Big[ -\frac{\omega}{\gamma(k)}\Big ]\frac{E(k)\, dk}{\gamma(k)}\,,
\end{equation}
 derived in \Refn{2013-ELB}  for the turbulence of Newtonian fluids in the framework of the Navier-Stokes equation in the Belinicher-L'vov sweeping-free representation\cite{1987-BL,1995-LP}.   In \Eq{bridge},   the turnover frequency $\gamma(k)$ of turbulent fluctuations of size $\sim \pi/k$ (referred below as $k$-eddies) can be estimated as usual:

\begin{align}\label{gamma} 
\gamma(k) &=C_\gamma  \sqrt {k^3\, E(k)}\ . \end{align}
Here $C_\gamma$ is a dimensionless constants of the order of unity.

To clarify the physical mechanisms behind the bridge  \Eq{bridge}, we consider the developed turbulence as consisting of an ensemble of $k$-eddies of the energy density  $E(k)$. Each such $k$-eddy is a random motion  with velocities $v_k(\tau)$ in the reference frame comoving with the center of the eddy.  To describe its statistical behavior, we introduce a new object   --  partial correlation function of $k$-eddies 
\begin{subequations}\label{11}
	\begin{equation}\label{C}
	\widetilde{\C C}(\tau,k)=  \frac{\< v_k(0) v_k^*(\tau)\>_k}  { \< |v_k(0)|^2  \>_k }\,,
	\end{equation}
	where $\< \dots \>_k$ denotes averaging of ensemble of  $k$-eddies with a given value of $k$. By definition, $\widetilde{\C C}(0,k)=1$. Using \Eq{C}, we introduce a  decorrelation time  of $k$-eddies in the Lagrangian framework $\tau_k$, defined such that 
	$\widetilde{\C C}(\tau_k,k)\simeq \frac12$. In developed turbulence, $\tau_k$ is about the life-time of $k$-eddies or their turnover time.  Below we sometimes use the frequency $\gamma_k\=1/\tau_k$  instead of $\tau_k$.

 The Fourier transform  of $\widetilde{\C C}(\tau,k)$ 
	\begin{equation}\label{11A}
	\widetilde{\C E}(\omega, k)= \int \limits_{-\infty}^{\infty} \exp(i\omega \tau)\widetilde{\C C}(\tau,k)d\tau\,,
	\end{equation}
	 according to the bridge \Eq{7a}, is nothing  but the frequency power spectrum of the considered $k$-eddy.
	The corresponding inverse Fourier transform 
	\begin{equation}\label{11B}
	\widetilde{\C C}(\tau, k)= \int \limits_{-\infty}^{\infty} \exp(-i\omega \tau)\widetilde{\C E}(\omega,k)\frac {d \omega}{2\pi}   
	\end{equation}
	dictates the normalization of $\widetilde{\C E}(\omega, k)$
	\begin{equation}\label{11C}
	\int \limits_{-\infty}^{\infty}  \widetilde{\C E}(\omega,k)\frac {d \omega}{2\pi}=\widetilde{\C C}(0, k)=    1\ .
	\end{equation}\end{subequations}
It can be shown\,\cite{1995-LP} that  for $\omega\tau_k\gg 1$, the  contribution of $k$-eddies  to $\widetilde{\C E}(\omega,k)$ decays much faster than $1/\omega^2$.

Assuming for simplicity the exponential decay,  it was suggested\cite{2013-ELB} that 
$\~{\C E}(\omega,k)\propto \exp(-\omega\tau_k)=\exp(-\omega/\gamma_k)$.   This assumption, together with the normalization\,\eqref{11C}, results in the model expression for the contribution of $k$-eddies to the Lagrangian frequency energy spectrum:  
\begin{subequations}\label{12}\begin{equation}\label{12A}
	\widetilde{\C E}(\omega,k)=\frac{2\pi}{ \gamma_k} \exp\Big(-\frac{\omega} {\gamma_k}\Big)\ . 
	\end{equation}
	To sum up contributions of all $k$-eddies to the frequency spectrum, we have to integrate the $k$-eddy contribution $\widetilde{\C E}(\omega,k)$ over $k$  with the weight $E(k)$, i.e. the energy distribution of $k$-eddies:
	\begin{equation}\label{12B}
	\C E (\omega)= \int 	\widetilde{\C E}(\omega,k) E(k) dk\ .
	\end{equation}\end{subequations}
 Combining \Eq{12A} and \eqref{12B}, we finally get the bridge\,\Eq{bridge} for $\C E(\omega)$. This equation   satisfies the exact general requirement: the total energy density per unit mass does not depend on the representation
\begin{equation}\label{sum-rule}
\overline E=\int E(k) \, dk=\int \C E(\omega) \, \frac{d\omega}{2\pi}=C(0) \ .
\end{equation}
The Eulerian-Lagrangian one-way bridge \Eq{bridge} allows us to find the Lagrangian (frequency) kinetic energy spectrum $\C E(\omega)$, for a given Eulerian energy spectrum $E(k)$. It is important to note that \Eq{bridge}
is  not limited by either the inertial interval of scales, or by the requirement of large Reynolds numbers.

\subsection{Bridge equations for model spectra}
  To illustrate the bridge  \Eqs{4b}, we suggest  a set of model expressions for  the Eulerian energy spectra, based on the Kolmogorov scaling: 
 \begin{align}\begin{split}\label{mod}
 E_1(q,m) &= \frac{q^m}{1+q^{m+5/3}}\,, \\
 E_2(q,m)&= \frac{10+q^4}{0.1+q^4}E_1(q,m) \ .
 \end{split}\end{align}
 plotted in \Fig{f:1}(a) as functions of the dimensionless wavenumber $q$.  These spectra have the K41 scaling  $E(q,m)\propto q^{-5/3}$ for $q>1$. To have large enough inertial interval, we choose for concreteness $q\sb{max}=1024$ and take $E_1(q,m)=E_2(q,m)=0$ for $q>q\sb{max}$.  In the range of small $q$, the model functions $E(q,m)$ demonstrate a variety of  possible behaviors:  $E_1(q,0)$ (the orange line) is approaching a plateau, while   spectra $E_1(q,2)$, $E_2(q,2)$ and $E_2(q,4)$ (the red, blue and green lines respectively)  decay for $q \ll 1$, going through a maximum for last two cases. 
 
 The normalized correlation functions 
 \begin{subequations}\label{norm}	
 	\begin{align}\begin{split}\label{6a}
 	&	\widehat{C}_1(y,m)\=  C_1(y,m)/C_n(0,m)\,, \\
 	&	\widehat{C}_2(y,m)\=  C_2(y,m)/C_n(0,m)\,, 
 	\end{split}	\end{align}
 	found from the corresponding spectra $E_1(q,m),E_2(q,m)$ with the help of  \Eqs{4A} and \eqref{4B},  are shown in \Fig{f:1}(b) as a function of a dimensionless coordinate $y$ with the same color code as in \Fig{f:1}(a). Correlation functions $\widehat C_1(y,0)$  and $\widehat C_1(y,2)$ (originated from the spectra which for small $q$ lie below K41 asymptote)  have asymptotic behavior  $\widehat{C}_n(y)= 1- y^{2/3}$, shown in \Fig{f:1}(b) as black dashed line,  over relatively large interval $y<1$.  On the other hand, $\widehat C_2(y,2)$  and $\widehat C_2(y,4)$ deviate from it  everywhere except for a narrow range $y<0.1$, not visible in the linear scale. As expected, all correlation functions $\widehat  C(y,m)$ vanish for large $y$ (in our case for $y>10$) but in different ways: monotonically [e.g. $\widehat C_1(y,0)$], crossing zero once  or twice. 
 	The typical scaling ranges are seen for the  normalized structure function 
 	\begin{equation}\label{6b}
 	\widehat{S}_n(y)\=\frac{S(y)}{2 C(0)}= 1 -\widehat{C}_n(y)\,, 
 	\end{equation}		
 \end{subequations}
 plotted in \Fig{f:1}(c). Indeed, for small $y<0.003$
 the viscous behavior $\widehat S(y,m)\propto y^2$ (shown by black dashed lines) is observed. This scaling  is expected whenever the energy spectrum decays faster than $q^{-3}$, including the sharp cutoff of the energy spectra for $q> q\sb{max}=1024$ in our model. For larger $y$, the  structure functions follow closely the K41 scaling $\widehat S(y,m)\propto y^{2/3}$, shown in \Fig{f:1}(c) as dot-dashed lines.  At large scales,  all structure functions demonstrate a tendency to approach plateau $\widehat{S}(y,m)\to 1$, as expected.

 Note that substituting $\widehat{C}(y,m)$  into \Eqs{4D}, we obtain  again the initial spectra $E(q,m)$, shown in \Fig{f:1}(a). We conclude that using the bridges \Eqs{4b} with one of the functions $E(q)$, $C(y)$ or $S(y)$, we can compute the rest  of them.  This means that all three considered characteristics of turbulence,  the energy spectra, the correlation and the   structure function, contain the same information about the second-order  statistics   of turbulence. However, they stress different aspects of the turbulence statistics: the small scale characteristics are highlighted  by $\widehat S(y)$ [\Fig{f:1}(c)], the large scale behavior is exposed by $\widehat C(y)$ [\Fig{f:1}(b)],  while  the energy distribution in wavenumber space is described by $E(q)$  [\Fig{f:1}(a)].

 \begin{figure*}[htp]
 	\includegraphics[scale=0.39]{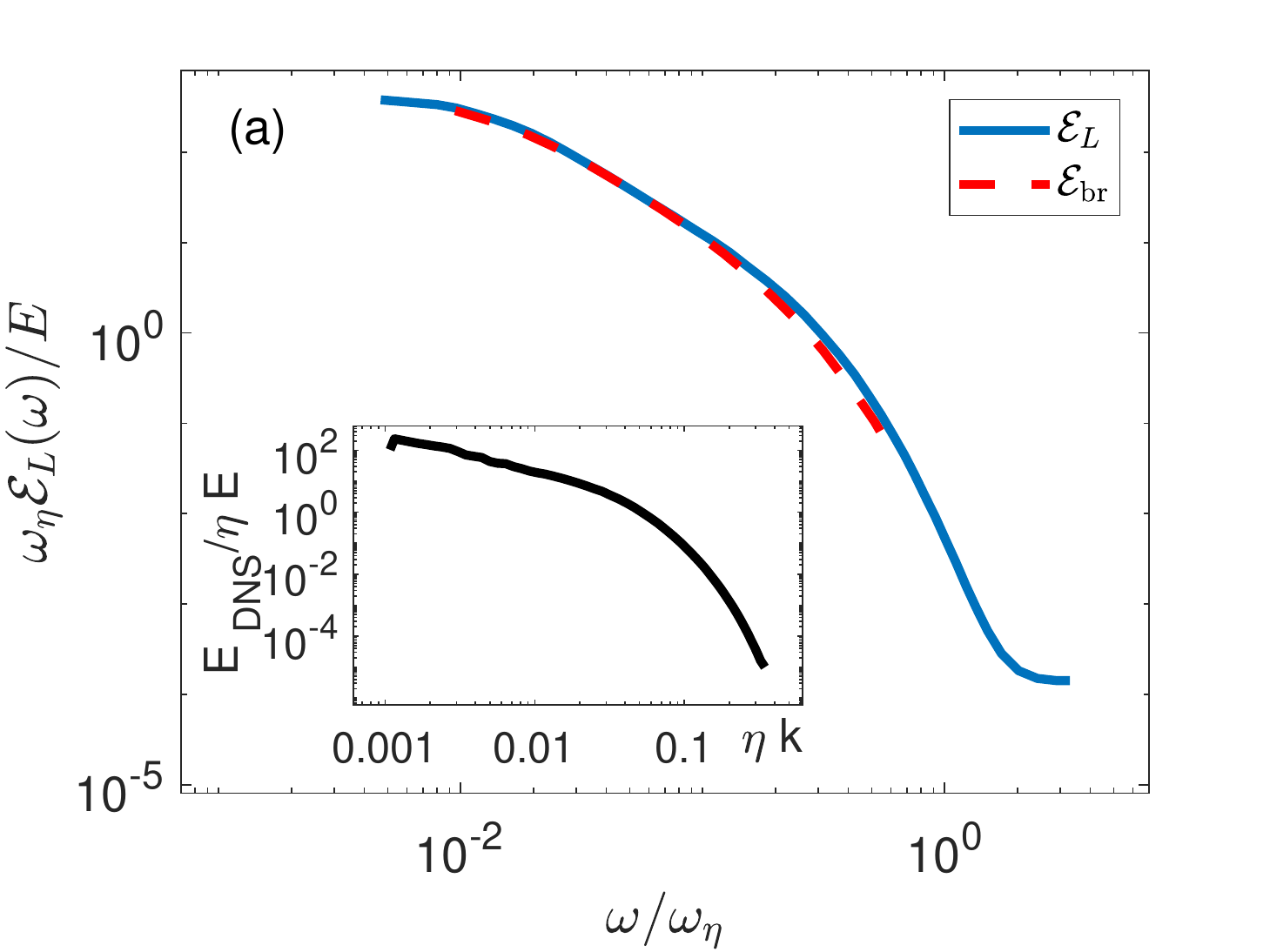} 
 	\includegraphics[scale=0.39]{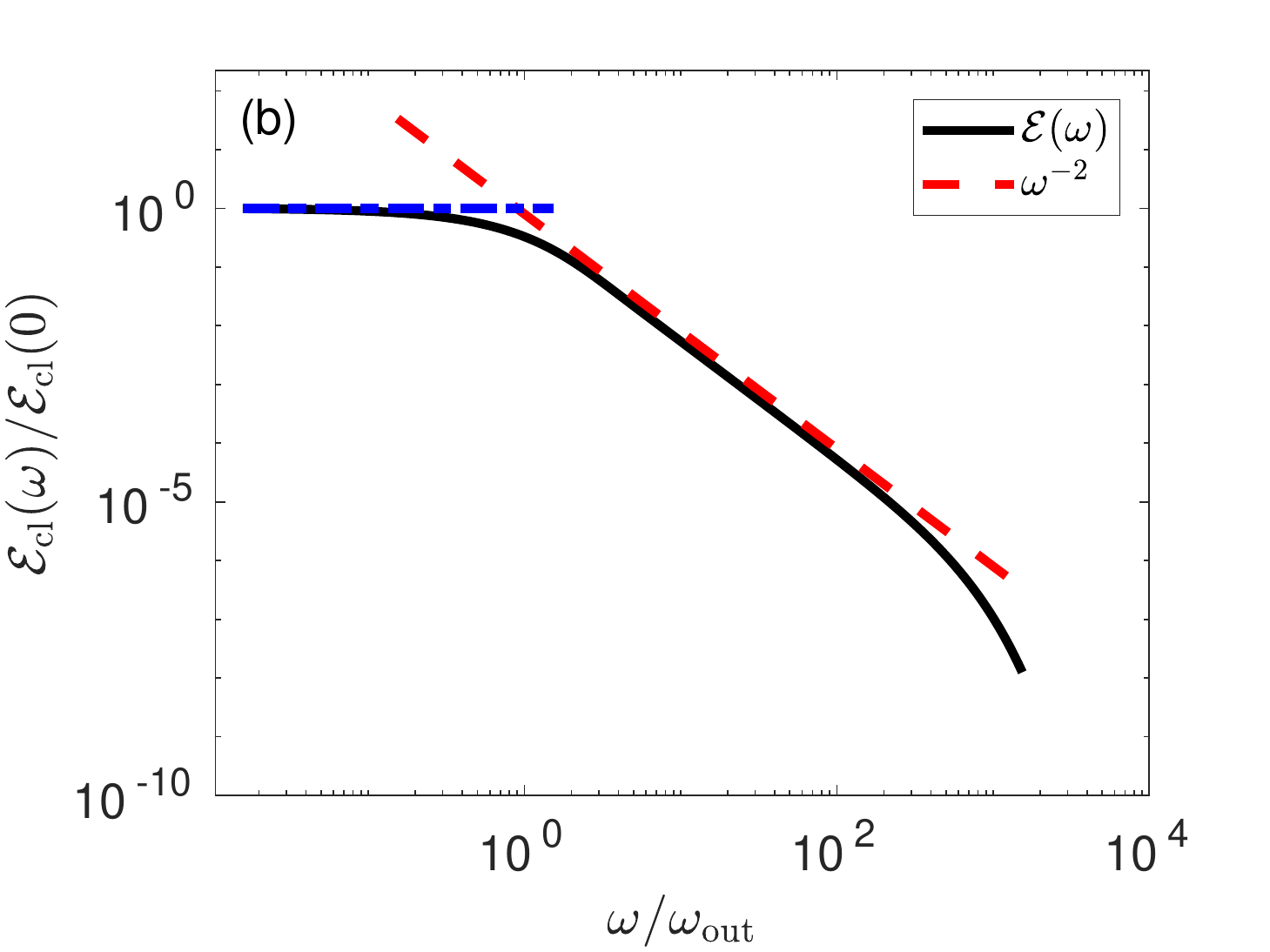}
 	\includegraphics[scale=0.39]{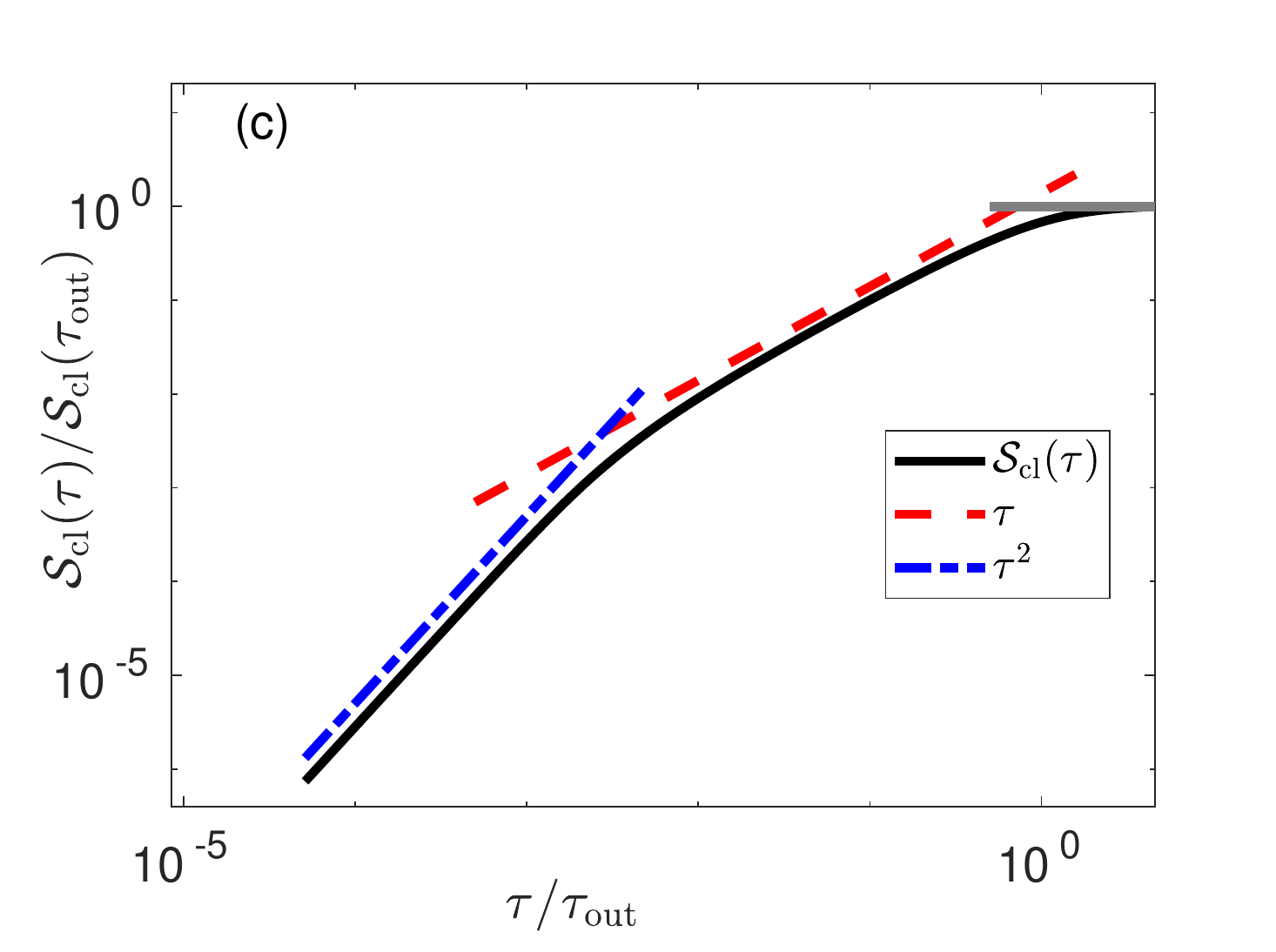} 
 	\caption{\label{f:smod}  Eulerian and Lagrangian statistics in classical turbulence.
 		(a)	Energy spectra  in DNS\,\cite{2013-ELB} with  $N=1024^3$ and Re$_\lambda=240$.  The Lagrangian energy spectrum $\C E(\omega)$ (solid line) in comparison with the spectrum $\C E\sb{br}$ (dashed line) calculated from the bridge \Eq{bridge} by integrating $E\Sb{DNS}(k)$, shown in the inset. The data of the dashed line ($0.01 \leq \omega/\omega_\eta \leq 0.6$) are limited by the	available DNS data.
 		(b) The Lagrangian  energy spectrum $\C E\sb{cl} (\omega)$, reconstructed using the bridge\,\eqref{bridge} from  K41 Eulerian spectrum \Eq{Emod} in the interval $0.01< k < 100$ and zero elsewhere.  Blue dot-dashed line shows the asymptote  $\displaystyle \lim_{\omega\to 0}\C E (\omega)$, red dashed line denotes K41 scaling in the inertial range $\C E (\omega)\propto\omega^{-2}$. (c) The Lagrangian structure function \Eq{7c} calculated using the spectrum $\C E(\omega)$, shown in (b).   Blue dot-dashed line shows viscous asymptotic behavior $\C S(\tau)\propto \tau^2$, red dashed line denotes K41 inertial range scaling $\C S(\tau)\propto \tau $ and gray horizontal line marks large $\tau$ limit $\C S(\tau)=$const.
 	}	
 \end{figure*}

 The   Eulerian-Lagrangian bridge equation\,\eqref{bridge} was verified in \Refn{2013-ELB} by DNS of Navier Stokes equations for stationary isotropic developed turbulence with 1024$^3$ grid points, as is illustrated in \Fig{f:smod}(a). In these simulations,  Re$_\lambda=240$ was reached, which  allowed the extend of an inertial interval of about 20 with K41 scaling $E\Sb{DNS}(k)\propto k^{-5/3}$ and well resolved viscous subrange, see the inset of \Fig{f:smod}(a). In this turbulent flow, the Lagrangian trajectories of $5\times 10^5$ fluid points were computed during about 20 large-eddy turnover times.
The instantaneous  Lagrangian velocity of the fluid point, required to integrate its trajectory, is computed using the fourth-order
three-dimensional Hermite interpolation  from the fluid velocity at the eight Eulerian grid nodes surrounding the fluid point. \Fig{f:smod}(a) shows excellent agreement directly between the Lagrangian spectrum $\C E\Sb{DNS}(\omega)$  found in the DNS (black solid line) and  the
spectrum $\C E\sb{br}(\omega)$ (red dashed line), calculated from the Eulerian spectrum $E\sb{DNS}(k)$ using the bridge. 

 To illustrate the bridge equations \Eqs{bridge} and \eqref{7c}, we use a simple model spectrum that has a form 

\begin{equation}\label{Emod}
E\sb{mod}(k)=k^{-5/3}   \,,  \quad \gamma(k)=k^{2/3}\,,
\end{equation}
in the inertial interval between $k\sb{min}=0.01\,$,  $k\sb{max}=100\,$ and is zero elsewhere.  The resulting Lagrangian spectrum $\C E\sb{cl}(\omega)$ is shown in  \Fig{f:smod}(b).   As expected, it has the  K41 scaling\,\eqref{K41},  $\C E(\omega)\propto 1/\omega^2$  in the inertial interval of frequencies  (shown by the red dashed line), in our  case between $\omega\sb{min}=0.2$ and $\omega\sb{max}=20$.  It approaches some constant value in the limit of $k\to 0$ and is experiencing an exponentially fast decay in the viscous sub-range.   This behavior   may be reproduced by the   following interpolation formula:
\begin{eqnarray}\label{17}
  \C  E(\omega) = \frac 2 \pi \,\overline E \, \omega\sb{min}\Big [ \frac 1{(\omega+\omega\sb{min})^2} -  \frac 1{(\omega\sb{max}+\omega\sb{min})^2}  \Big]
\end{eqnarray}
for all frequencies   $\omega<\omega\sb {max}$.  The only difference is a sharp cutoff at $\omega=\omega\sb {min}$ except of a smooth exponential decay.
  For $\omega\sb{max}\gg \omega\sb{min}$, \Eq{12} satisfies the  same frequency sum-rule\,\eqref{sum-rule} as the numerical spectrum $\C E\sb{cl}(\omega)$.

The corresponding  Lagrangian  structure function $\C S\sb{cl}(\tau)$ calculated using \Eq{7c}, is shown in \Fig{f:smod}(c). As expected, it has the K41 scaling $\C S(\tau)\propto \tau$ in the inertial interval of scales, shown  by red dashed line, and the viscous behavior $\C S(\tau)\propto \tau^2$, shown by blue dot-dashed line. At very large times, $\C S(\tau)$  approaches a  constant value.

 \section{\label{s:2}  Experiment    and data analysis}
 %\vskip -.1 cm

 \begin{table*}%[!ht]	
 	\centering
 	\caption{Key parameters of the grid turbulence in He II.}
 	\label{table:KeyP}
 	\begin{center}
 		\setlength{\tabcolsep}{4mm}%{
 		\begin{threeparttable}
 			\begin{tabular}{m{90pt}m{115pt}ccc}
 				\hline\hline
 				\specialrule{0em}{2pt}{2pt}
 				Parameters& Expression$^{\text{a},\text{b}}$ & 1.65 K &1.95 K &2.12 K\\
 				\hline				
 				\specialrule{0em}{2pt}{2pt}			
 				Vortex line density &$\C L$, (cm$^{-2}$) &2.1$\times 10^4$ &$3.7\times 10^4$ &$1.9\times 10^4$\\ 
 				\specialrule{0em}{2pt}{2pt}
 				Intervortex distance &$\ell=1/\sqrt{\C L}$, (mm) &0.07 &0.05 &0.07\\
 				\specialrule{0em}{2pt}{2pt}
 				\multirow{2}{90pt}{The crossover wave number} &\multirow{2}{115pt}{$k\sb{cr}= 2\pi/\ell$, (mm$^{-1}$)} &  \multirow{2}{20pt}{\centering 89.7} &\multirow{2}{20pt}{\centering 125.6} &\multirow{2}{20pt}{\centering 89.7}\\[10pt]
 				\specialrule{0em}{2pt}{2pt}
 				\multirow{2}{90pt}{The outer scale of turbulence\cite{WG1}} &$L\sb{out}$, (mm) &3 &2 &3\\
 				\cmidrule(){2-5}
 				&$k\sb{out} \simeq 2\pi/L\sb{out}$, (mm$^{-1}$) &2.1 &3.1 &2.1\\
 				\specialrule{0em}{2pt}{2pt}
 				\multirow{3}{90pt}{Mean density of the kinetic energy per unit mass} &\multirow{3}{115pt}{$\overline E \equiv \langle  |\B u(\B r,t)|^2\rangle _{\B r}$, (mm$^2$/s$^2$)} &\multirow{3}{20pt}{\centering 13.2} &\multirow{3}{20pt}{\centering 16.6} &\multirow{3}{20pt}{\centering 9.9}\\[21pt]
 				\specialrule{0em}{2pt}{2pt}				
 				\multirow{2}{90pt}{RMS of the turbulent velocity} &\multirow{2}{115pt}{$v\Sb T= \sqrt {\overline E}$, (mm/s)} &\multirow{2}{20pt}{\centering 3.6} &\multirow{2}{20pt}{\centering 4.1 }&\multirow{2}{20pt}{\centering 3.1} \\[10pt]
 				\specialrule{0em}{2pt}{2pt}
 				\multirow{2}{90pt}{Outer-scale turnover frequency} &\multirow{2}{115pt}{$\omega\sb{out}  \simeq 2\pi  v \Sb T / L\sb{out}$, (s$^{-1}$)} &\multirow{2}{20pt}{\centering 7.5} &\multirow{2}{20pt}{\centering 12.9} &\multirow{2}{20pt}{\centering 6.5}\\[10pt]
 				\specialrule{0em}{2pt}{2pt}
 				\multirow{3}{90pt}{Turnover frequency of the smallest eddies of scale $\ell$} &\multirow{3}{115pt}{$\omega_\ell   \simeq  \omega\sb{out} (L\sb{out} /\ell )^{2/3}$, (s$^{-1}$)}  & \multirow{3}{20pt}{\centering 92.3} &\multirow{3}{20pt}{\centering 150.6} &\multirow{3}{20pt}{\centering 79.5}\\[23pt]
 				\specialrule{0em}{2pt}{2pt}
 				\multirow{3}{90pt}{Energy dissipation rate\cite{DXu-2013-ETFS} $\epsilon$} &\multirow{3}{115pt}{$\epsilon = \nu\langle4(\frac{\partial v_x}{\partial x})^2+4(\frac{\partial v_z}{\partial z})^2+3(\frac{\partial v_x}{\partial z})^2+3(\frac{\partial v_z}{\partial x})^2+4(\frac{\partial v_x}{\partial x}\frac{v_z}{\partial z})+6(\frac{\partial v_x}{\partial z}\frac{\partial v_z}{\partial x})\rangle$, (mm$^2$/s$^3$)}  & \multirow{3}{20pt}{\centering 21.2} &\multirow{3}{20pt}{\centering 43.7} &\multirow{3}{20pt}{\centering 11.8}\\[35pt]
 				\specialrule{0em}{2pt}{2pt}
 				\multirow{3}{90pt}{Kolmogorov microscale $\eta$} &\multirow{3}{115pt}{$\eta = (\frac{\nu^3}{\epsilon})^{\frac{1}{4}}$, ($\mu$m)}  & \multirow{3}{20pt}{\centering 17.4} &\multirow{3}{20pt}{\centering 11.9} &\multirow{3}{20pt}{\centering 24.5}\\[15pt]
 				\specialrule{0em}{2pt}{2pt}
 				\multirow{3}{90pt}{Kolmogorov time scale $\tau_\eta$} &\multirow{3}{115pt}{$\tau_\eta = (\frac{\nu}{\epsilon})^{\frac{1}{2}}$, (ms)}  & \multirow{3}{20pt}{\centering 33.8} &\multirow{3}{20pt}{\centering 14.8} &\multirow{3}{20pt}{\centering 45.3}\\[15pt]			
 				\specialrule{0em}{2pt}{2pt}
 				\multirow{3}{90pt}{Stokes time $\tau_\text{s}$} &\multirow{3}{115pt}{$\tau_\text{s} = \frac{\rho_pd_\text{p}^2}{18\mu}$, (ms)}  & \multirow{3}{20pt}{\centering 0.22} &\multirow{3}{20pt}{\centering 0.20} &\multirow{3}{20pt}{\centering 0.15}\\[20pt]
 				\specialrule{0em}{2pt}{2pt}
 				Stokes number{\cite{Jung-PRE-2008}} St\ &St = $\frac{\tau_{\text{s}}}{\tau_{\eta}}$ &0.007 &0.014&0.003\\
 				\hline\hline 
 			\end{tabular}
 			
 			\begin{tablenotes}
 				\footnotesize
 				\item[a] $\nu$ and $\mu$ denote the kinematic viscosity and the dynamic viscosity of He II\cite{1}.
 				\item[b] $d_p$ and $\rho_p$ denote the diameter and density\cite{TXu-2008} of the tracer particle.
 			\end{tablenotes}
 		\end{threeparttable}
 	\end{center}	
 \end{table*}

\subsection{\label{ss:exp}Description of the experiment and main parameters of the flow}

The experimental apparatus is described in detail in \Refn{WG1}. In particular, a transparent cast acrylic flow channel with a cross-section area of 1.6$\times$1.6 cm$^2$ and a length of 33 cm is immersed vertically in a He II bath whose temperature can be controlled by regulating the vapor pressure. A brass mesh grid with a spacing of 3 mm and 40\% solidity\cite{Mastracci-2018-RSI} is suspended in the flow channel by four stainless-steel thin wires at the four corners. These wires are connected to the drive shaft of a linear motor whose speed can be tuned in the range of 0.1 and 60 cm/s. In the current work, we used a fixed grid speed at 30 cm/s.

To probe the flow, we adopt the PTV method using solidified D$_2$ tracer particles with a mean diameter of about 5~$\mu$m~\cite{Mastracci-2018-RSI}. Due to their small sizes and hence small Stokes number in the normal fluid\cite{WG1}, these particles are entrained by the viscous normal-fluid flow. But when they are close to the quantized vortices in the superfluid, a Bernoulli pressure owing to the induced superfluid flow can push the particles toward the vortex cores, resulting in the trapping of the particles on the quantized vortices\cite{Mastracci-2018-PRF,Mastracci-2019-PRF,KBS-07,GK-19}.
The Stokes numbers at different temperatures are calculated and included in Table \ref{table:KeyP}.The Stokes number is calculated based on the ratio of the Stokes time to the Kolmogorov time scale\cite{Jung-PRE-2008}, which is in the range of 0.003-0.014 in all cases.

A continuous-wave laser sheet (thickness: 200~$\mu$m, height: 9 mm) passes through the center of the channel to illuminate the particles. We then pull the grid and use a high-speed camera (120 frames per second) to film the motion of the particles. This sampling time (i.e., 8.3 ms) is larger than the Stokes time but smaller than the Kolmogorov time in our experiment, which is desired for high fidelity velocity-field measurements (see \Refn{Tropea-2007}). Following the passage of the grid, we record the particle positions for a period of 0.28 s (i.e., 34 images) for every 2 s. A modified feature-point tracking routinehl\cite{Mastracci-2018-RSI} is adopted to extract the trajectories of the tracer particles from the sequence of images.
%\red{In the following, we define the coordinate system of the recorded images such that the streamwise direction corresponds to the positive $\B z$ axis,  while $ \B x$ axis  denotes the wall-normal direction.}
In the current work, we focus on analyzing the data obtained at 4 s ($T = 1.95$ K) and 6 s ($T = 1.65$ K, $T = 2.12$ K) following the passage of the grid. As discussed in \Refn{WG1}, the turbulence at these decay times appears to be reasonably homogeneous and isotropic, and its turbulence kinetic energy density is relatively high such that an inertial interval exists. We have also installed a pair of second-sound transducers for measuring the mean vortex line density $\C L(t)$ using a standard second-sound attenuation method\cite{Sherlock-1970-RSI}.

At a given temperature $T$, we normally repeat our measurement up to 10 times so that an ensemble statistical analysis of the particle trajectories can be performed. These 10 acquisitions, denoted as $A=1,\ 2 \dots 10$, each contains 34 consecutive images. The velocity of a particle can be calculated by dividing its displacement from one image to the next by the frame separation time. We only select the middle 24 images ($I=1,\, 2\, \dots 24$) for our velocity-field analysis. The velocities $\big \{ \B u(\B X)\big \}_{I,A}$ at the particle locations $\B X = (x,z)$ are determined, where $x$ and $z$ denote the horizontal (wall-normal) and the vertical (streamwise) coordinates, respectively

In \Table{table:KeyP}, we list some key parameters relevant to the flows that will be used in our result analysis. %\red{The small Stokes number implies that these particles can trace the fluid excellently\cite{Tropea-2007}. Also, our sampling time (8.3 ms), which is between the stokes time and the Kolmogorov time, meets the desire for high quality PTV measurements}.

  \subsection{\label{ss:Eul}2D velocity    on a periodic lattice from PTV data}
 \begin{figure}[!htbp]
	\includegraphics[width=0.6\columnwidth]{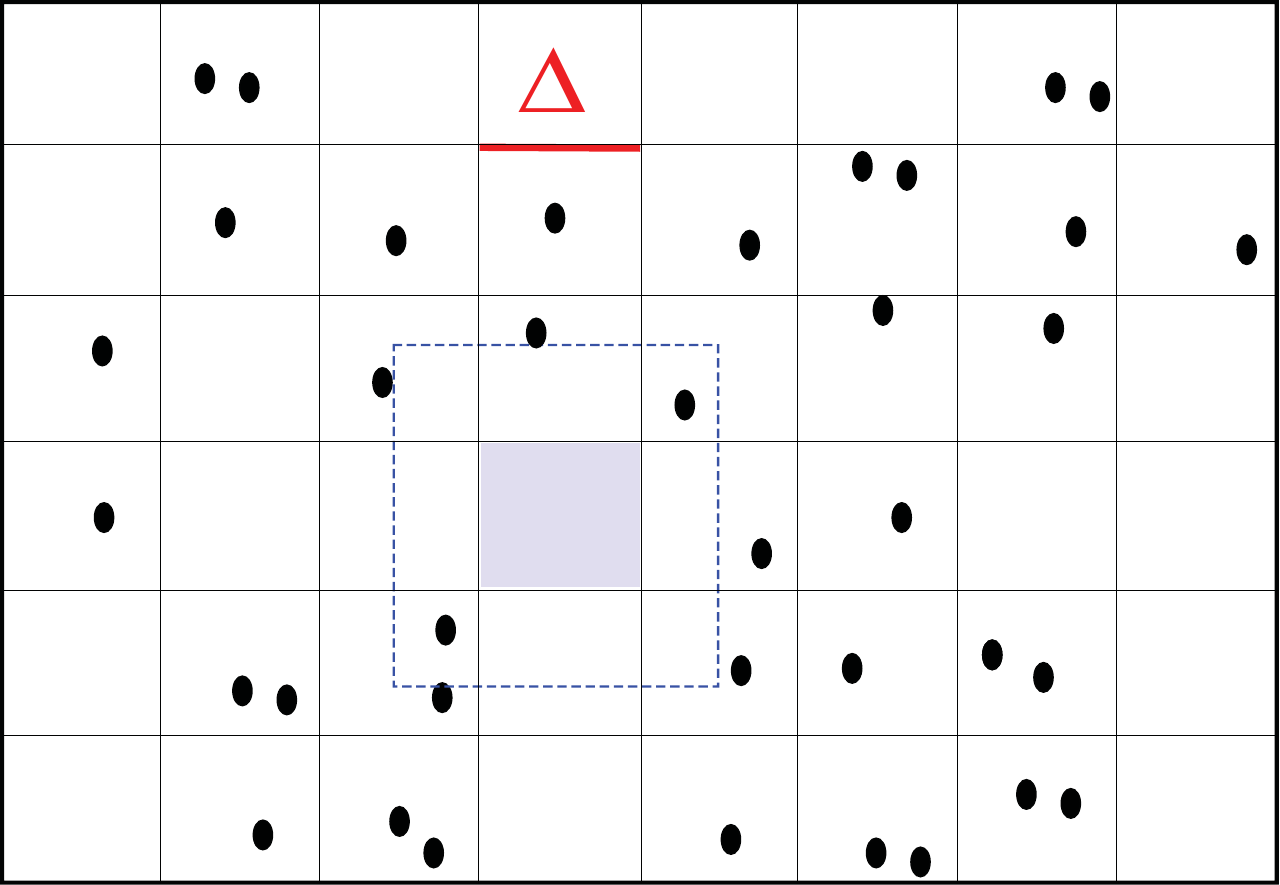}
	\caption{An illustration to the division of the observation
		area into cells to calculate the averaged velocity field.}
	\label{f:2}
\end{figure}
In order to analyze the Eulerian turbulence energy spectra, it is highly desired to generate a two-dimensional velocity field on a periodic lattice $\B R_{n,m}=\big \{ X_n=n\Delta, Z_m=m \Delta \big \}$. For this purpose, we first combine the velocity data $\B u(\B X)$ obtained from all 24 images in each acquisition into a single image. Then, we divide the combined image into square cells with $\Delta=0.02$ mm which is large enough to contain at least 1-2 data points in most of the cells, as shown in \Fig{f:2}. The velocity assigned to the center of each cell is calculated as the Gaussian-averaged velocity of particles inside the cell, $\B u(\B X)$, with the variance $\sigma\approx \Delta/2$ to guarantee that the Gaussian weight drops to near zero at the cell's edge. This procedure assumes that during the acquisition time of 0.2 s the velocity field does not change considerably so that these data describe a single instantaneous velocity field.
We have verified that averaging a smaller number of images (and hence shorter measurement time) does not alter significantly the large-scale velocity and spectra. 

  Occasionally, there may not be any particles that fall inside a particular cell. In this case, we increase the size of this cell by a factor of two as shown in \Fig{f:2}, and this process may be repeated until at least 1-2 particles fall in the enlarged cell so that the velocity at the cell center can be determined. The resulted velocity field $\big \{ \overline{\B u}(\B R_{n,m})\big \} $ is then Fourier transformed and the edge-related artifacts can be removed using known algorithms~\cite{artefact}. The ensemble-averaged Eulerian energy spectra can be derived based on these Fourier-transformed velocity fields.

 \subsection  {\label{ss:method}2D-energy spectra  and its 1D reductions  }
 
 Having obtained the velocity field $\big \{ \overline{\B u}(\B R_{n,m})\big \} $ on the periodic lattice $\B R_{n,m}$, we then perform the Fourier transform
 \begin{subequations}\label{13}
 \begin{align} \begin{split}\label{13a}
 \B u (k_x,k_z) = &\frac 1{N\, M}\sum_{n=0 }^{N-1}\sum_{m=0 }^{M-1}\overline{\B u}(\B R_{n,m})\\
 & \times \exp[ -i(k_x n+ k_z m)\Delta]\,,
 \end{split}\end{align}
 where $N$ and $M$ are the numbers of cells in $x$ and $z$ directions, respectively.
The 2D energy spectrum can be obtained as
 \begin{align}  \label{13b}
 F_\alpha(k_x,k_z)=  \langle |u_\alpha (k_x,k_z)|^2 \rangle \,,
 \end{align}\end{subequations}
 where $\alpha=x, \, z$ and $\langle \dots \rangle $ denotes an ensemble average over the the acquisitions $A$. Notice that \Eqs{13} are  the discrete 2D version of \Eqs{2}. 

In addition, we introduce the 1D  Fourier transforms
 \begin{align} \begin{split}
 \B u (k_x,z) = &\frac 1{N }\sum_{n=0 }^{N-1}\ \overline{\B u}(\B R_{n,m}) \exp[ -ik_x n)\Delta]\,,\\
 \B u (x,k_z) = &\frac 1{ M}\sum_{m=0 }^{M-1}\ \overline{\B u}(\B R_{n,m}) \exp[ -ik_z m)\Delta]\,,
 \end{split}\end{align}
 and 1D ``linear" energy spectra
 \begin{align} \label{elin}  \begin{split}
 E_\alpha^{\langle z \rangle}(k_x) & =  \langle |u_\alpha (k_x,z)|^2 \rangle_z \,,\\
 E_\alpha^{\langle x \rangle}(k_z) & =  \langle |u_\alpha ( x,k_z)|^2 \rangle_x \,,
 \end{split}\end{align}
where $ \langle \dots  \rangle_x$ (and $ \langle \dots  \rangle_z$) denotes averaging over $x$ (and the $z$) axis in addition to ensemble averaging over different acquisitions. Linear 1D spectra~\eqref{elin} are related to the 2D-spectra~\eqref{2} as follows:
 \begin{align}  \begin{split}\label{21}
 E_\alpha^{\langle z \rangle}(k_x) & =   \sum_{m=0}^{M-1} F_\alpha\Big(k_x ,\frac{2\pi m }{ M\,\Delta} \Big)\,, \\
 E_\alpha^{\langle x \rangle}(k_z) & =   \sum_{n=0}^{N-1} F_\alpha\Big(\frac{2\pi n }{N\,\Delta},k_z\Big)\ .
 \end{split}\end{align}

In order to examine possible anisotropy of the turbulence, we perform SO(2) decomposition to get 1D energy spectra. The decomposition is done by projecting 2D spectra\,\eqref{2} defined on the $(x,z)$-plane, on $0\sp{th}$ and  $p\sp{th}$  components of an orthonormal  basis, which is proportional to  $\exp( -i\, p\, \varphi) $:
 \begin{align}  \begin{split}\label{20}
 & \hskip - .1cm E_\alpha^{(0)}(k)  =  k \int\limits _0^{2\pi} F_\alpha(k\cos \varphi ,k\cos\varphi) d \varphi\,, \\
 & \hskip - .1cm  E_\alpha^{(p)}(k)  =  p \sqrt 2\, k \int\limits_0^{2\pi} F_\alpha(k\cos \varphi ,k\sin \varphi)\, \cos (p\varphi ) d \varphi\ .
 \end{split}\end{align}
 If we keep the lowest two components, the original 2D spectra\,\eqref{2} can be approximately expressed as:
 \begin{align}  \label{23}
 F_\alpha(k_x,k_z)=    E_\alpha^{(0)}(k)+ 2\sqrt 2 \sin (2\varphi ) E_\alpha^{(2)}(k)\ .
 \end{align}
For an isotropic turbulence, $E_\alpha^{(2)}(k)=0$. For flows with weak anisotropy, the strength of the anisotropy can be evaluated by the ratio $ E_\alpha^{(2)}(k) \big / E_\alpha^{(0)}(k)$.

 \begin{figure*}%[!ht]
	\begin{tabular}{ccc}
		
	 $T=1.65\,$K& $T=1.95\,$K &  $T=2.12\,$K \\
		\includegraphics[width=.7\columnwidth]{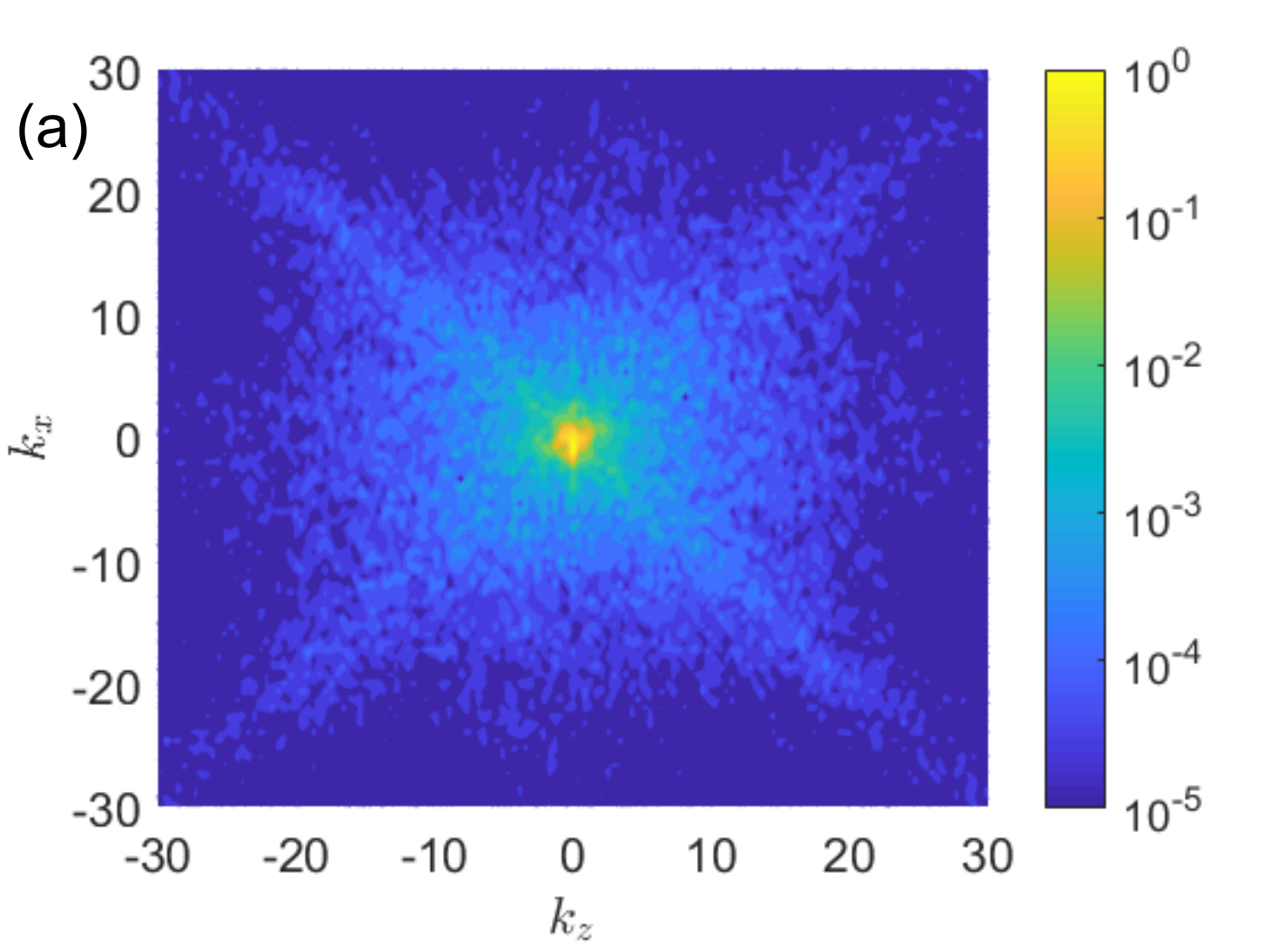} &
		\includegraphics[width=.7\columnwidth]{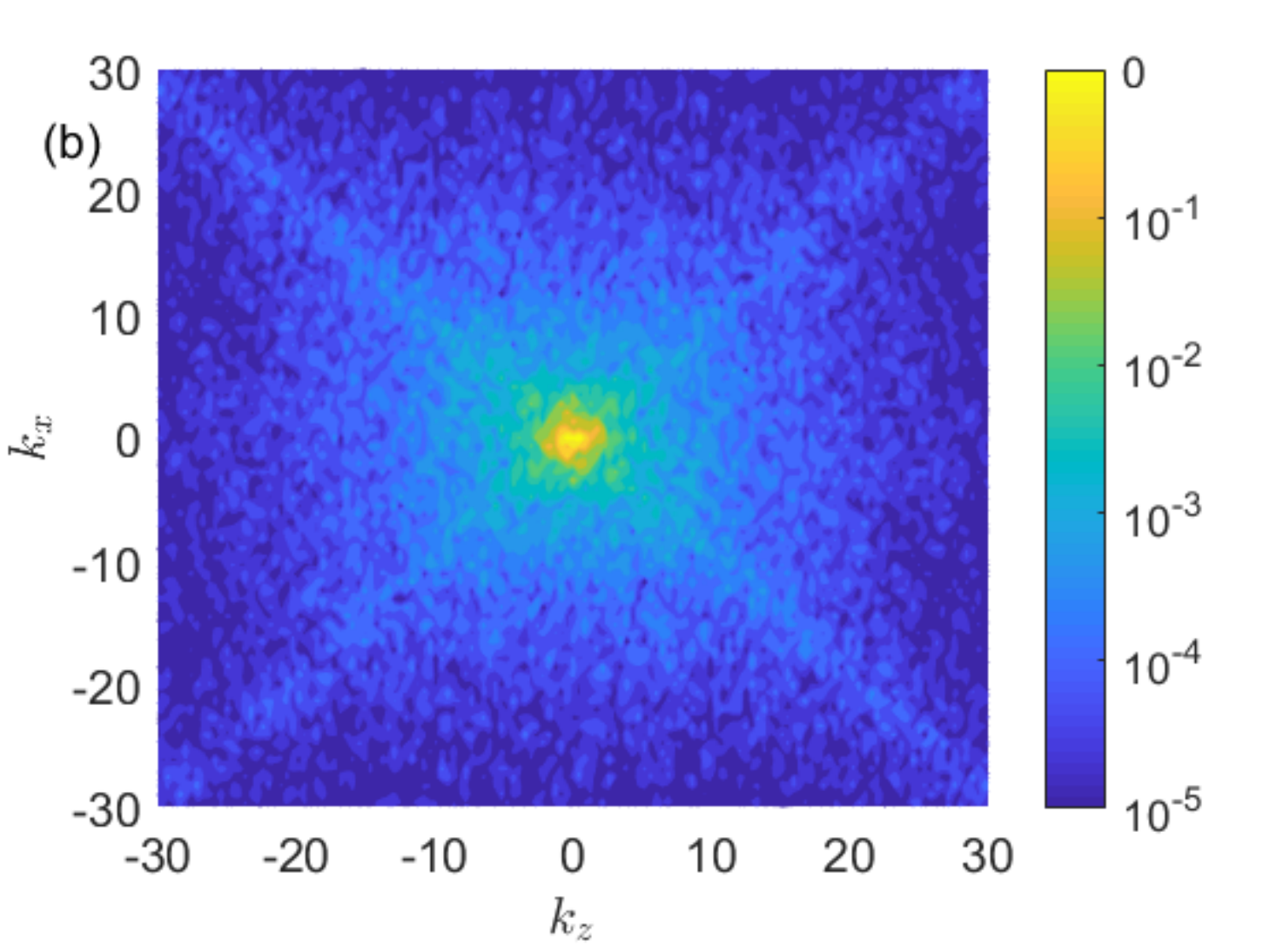} &
		\includegraphics[width=.7\columnwidth]{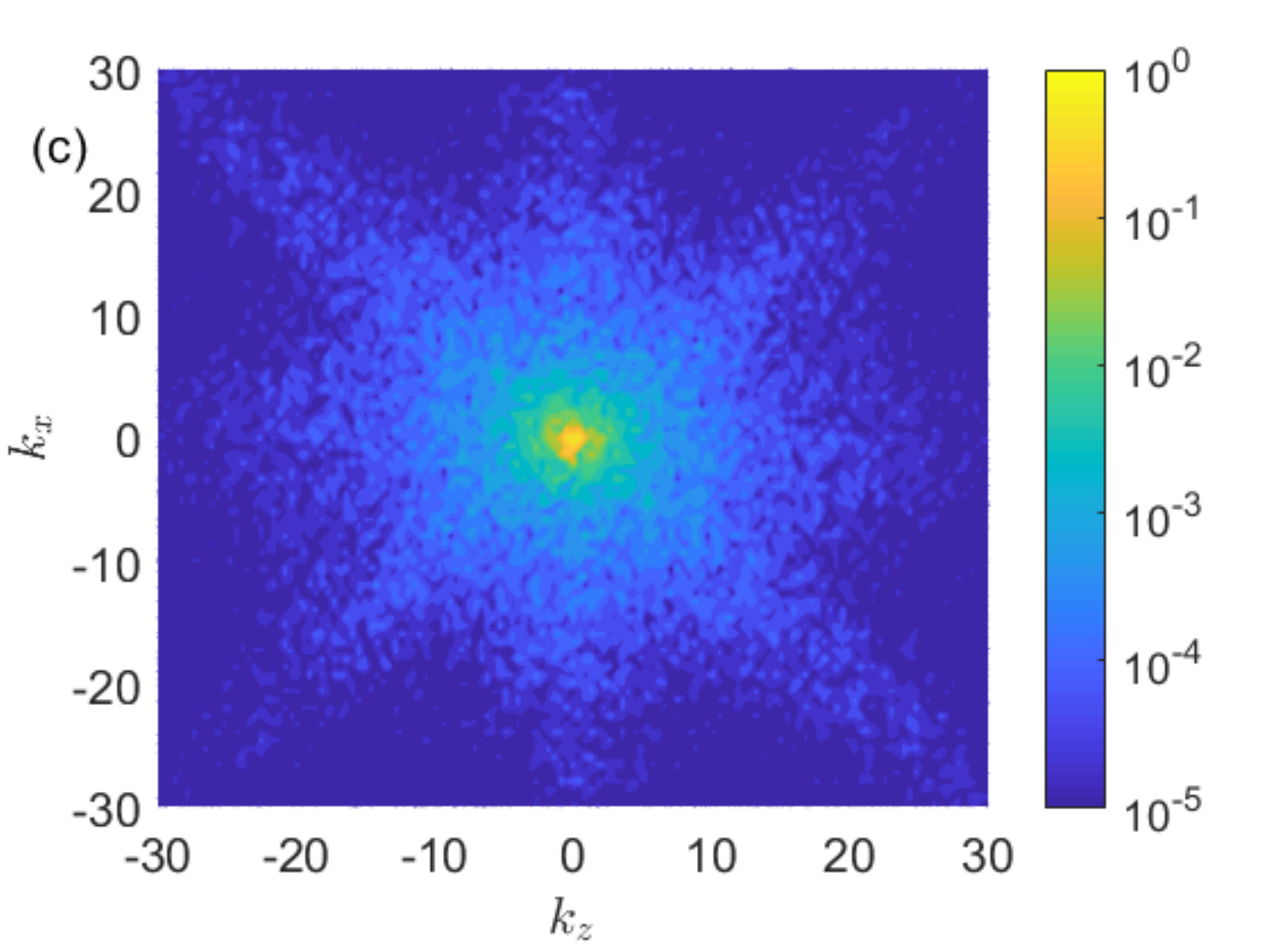}\\
	%	(d)& (e) & (f)\\
		\includegraphics[width=.7\columnwidth]{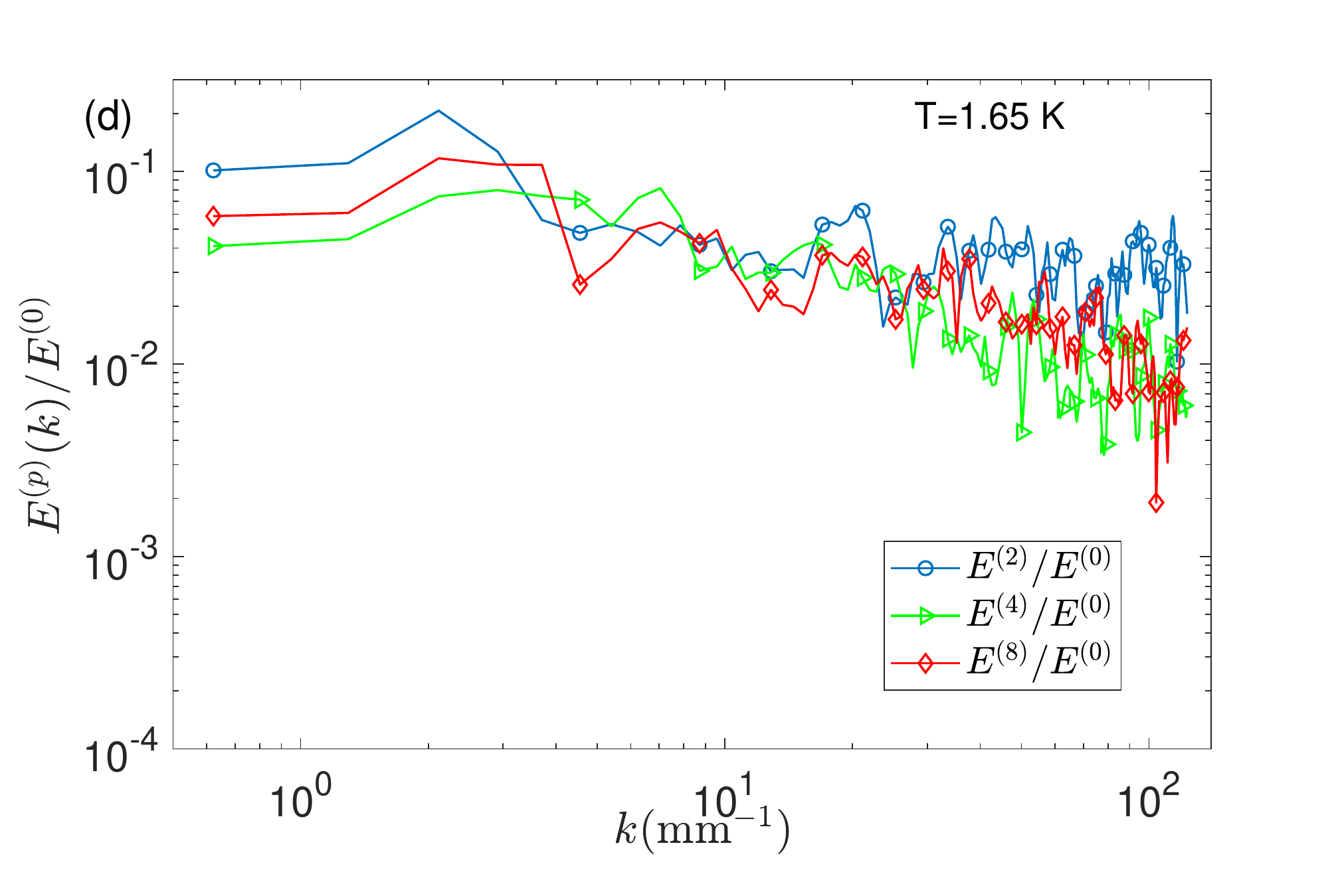} &
		\includegraphics[width=.7\columnwidth]{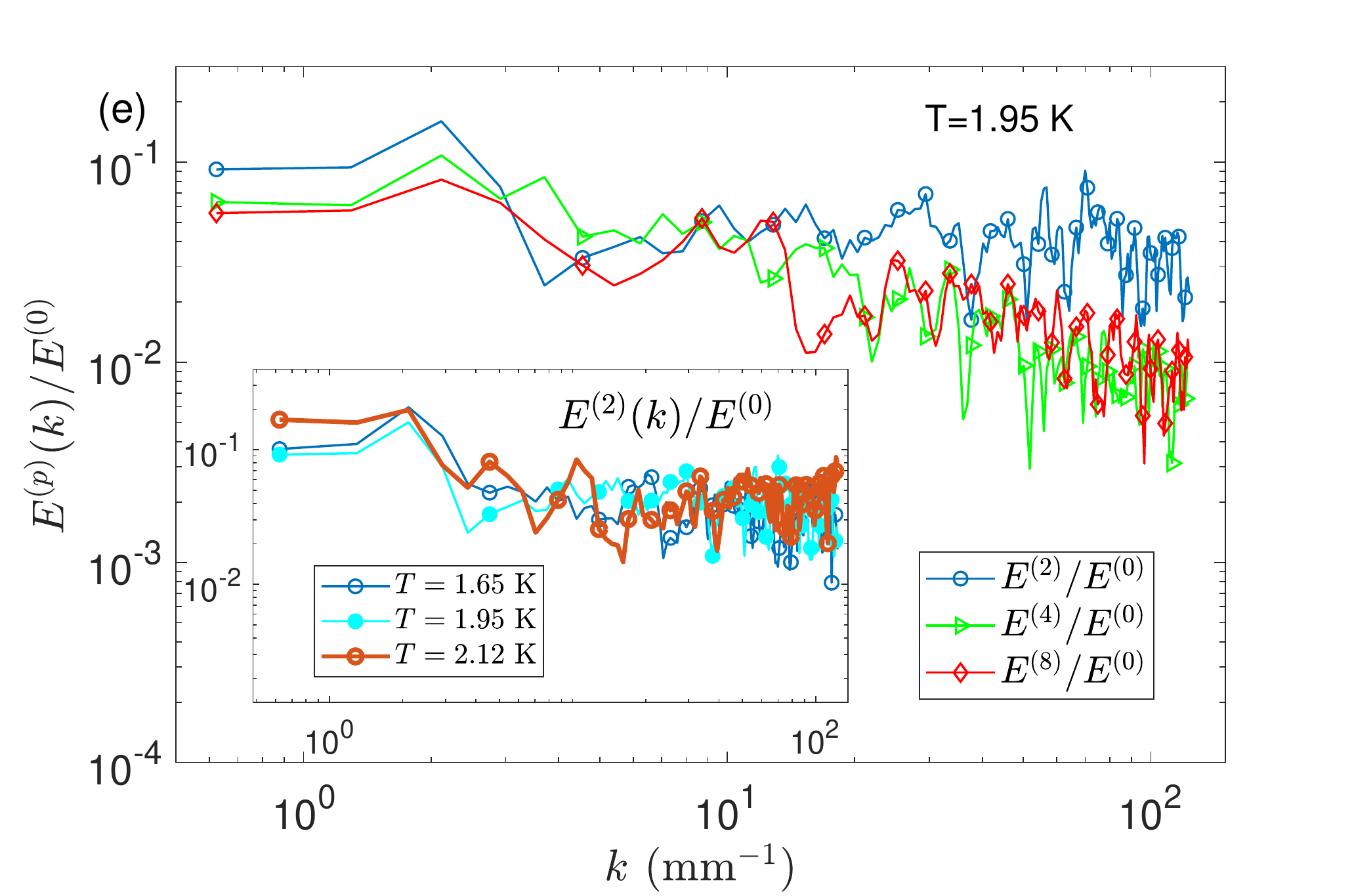} &
		\includegraphics[width=.7\columnwidth]{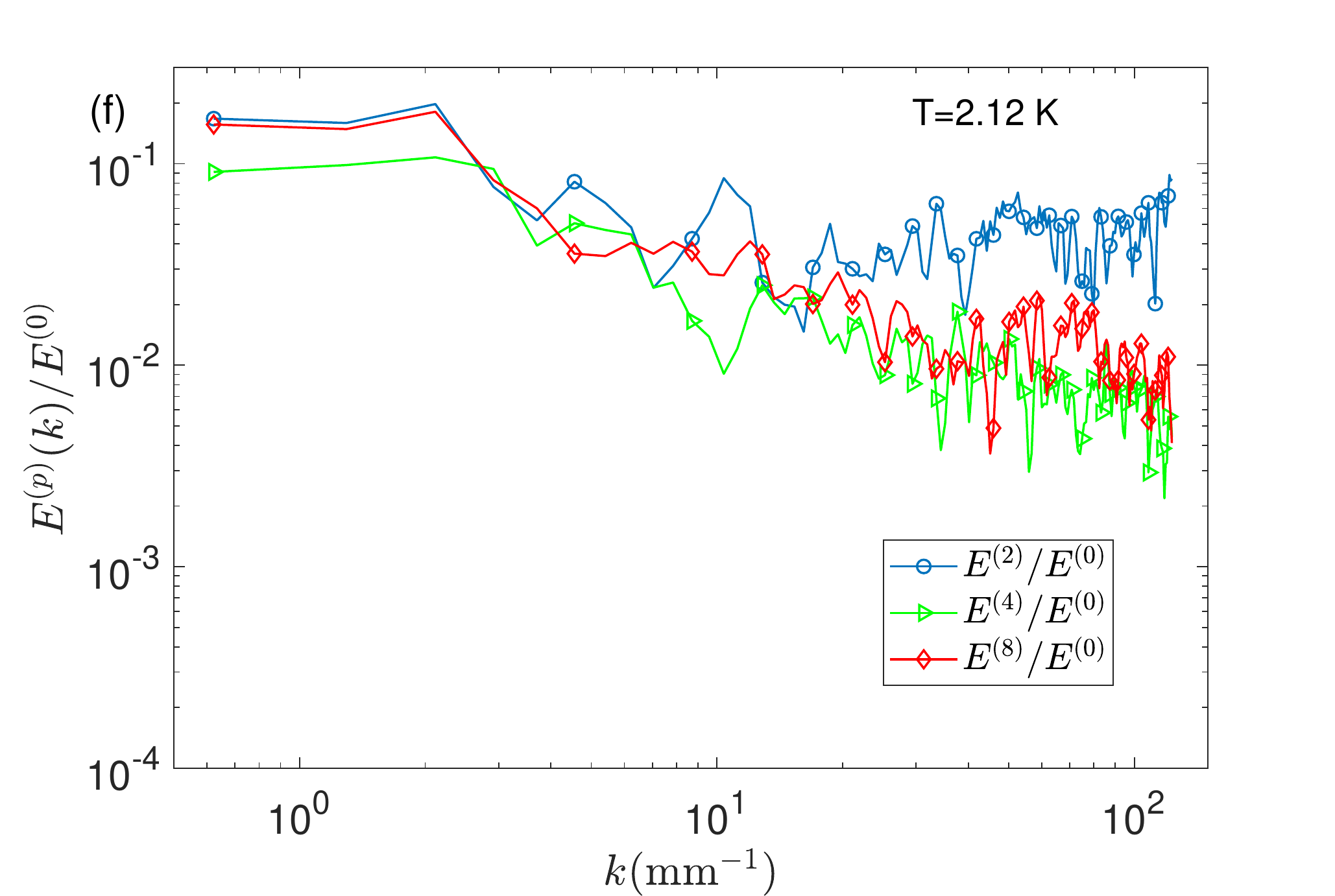} \\
	\end{tabular}
	\caption{\label{f:2d}  (a), (b) and (c):   2D energy spectra $F(k_x,k_z)$ for  $T=1.65\,$K,  $T=1.95\,$K and $T=2.12\,$K. Note logarithmic scale of the color code.  (d), (e) and (f):  The corresponding  ratios of SO(2), SO(4) and SO(8) decompositions of 2D energy spectra to its isotropic component,  $E^{(2)}(k)/E^{(0)}(k)$ (circles),   $E^{(4)}(k)/E^{(0)}(k)$ (triangles) and  $E^{(8)}(k)/E^{(0)}(k)$ (diamonds). The inset in (e) compares $E^{(2)}(k)/E^{(0)}(k)$  for different temperatures.
	}
\end{figure*}

\section{\label{s:spectra}Results and their discussion}

 \subsection{\label{ss:Euler} Eulerian statistics}
 \subsubsection{Eulerian energy spectra}
 Experimental results for 2D energy spectra $F(k_x,k_z)$ for T=$1.65\,\text{K,}\,1.95\,\text{K and}\,2.12\, \text{K}$,  are shown in \Fig{f:2d}(a), (b) and (c), respectively.  At all temperatures these spectra are nearly isotropic with small corrections.  In \Fig{f:2d}(d), (e) and (f), the ratios  $E^{(2)}(k)/E^{(0)}(k)$,   $E^{(4)}(k)/E^{(0)}(k)$ and  $E^{(8)}(k)/E^{(0)}(k)$ of the SO(2) decomposition components are shown for all temperatures. We see that except in regions of $k< 10\,$mm$^{-1}$, all anisotropic corrections are very small (below the level of 5\%) with respect of the    isotropic contribution $E^{(0)}(k)$.
We attribute the residual star-like structures in the 2D spectra to regions  of the velocity field originating from the cells with a small number of particles.

  \begin{figure*}[!ht]
  	\begin{tabular}{ccc}
  				
  		\includegraphics[width=.7\columnwidth]{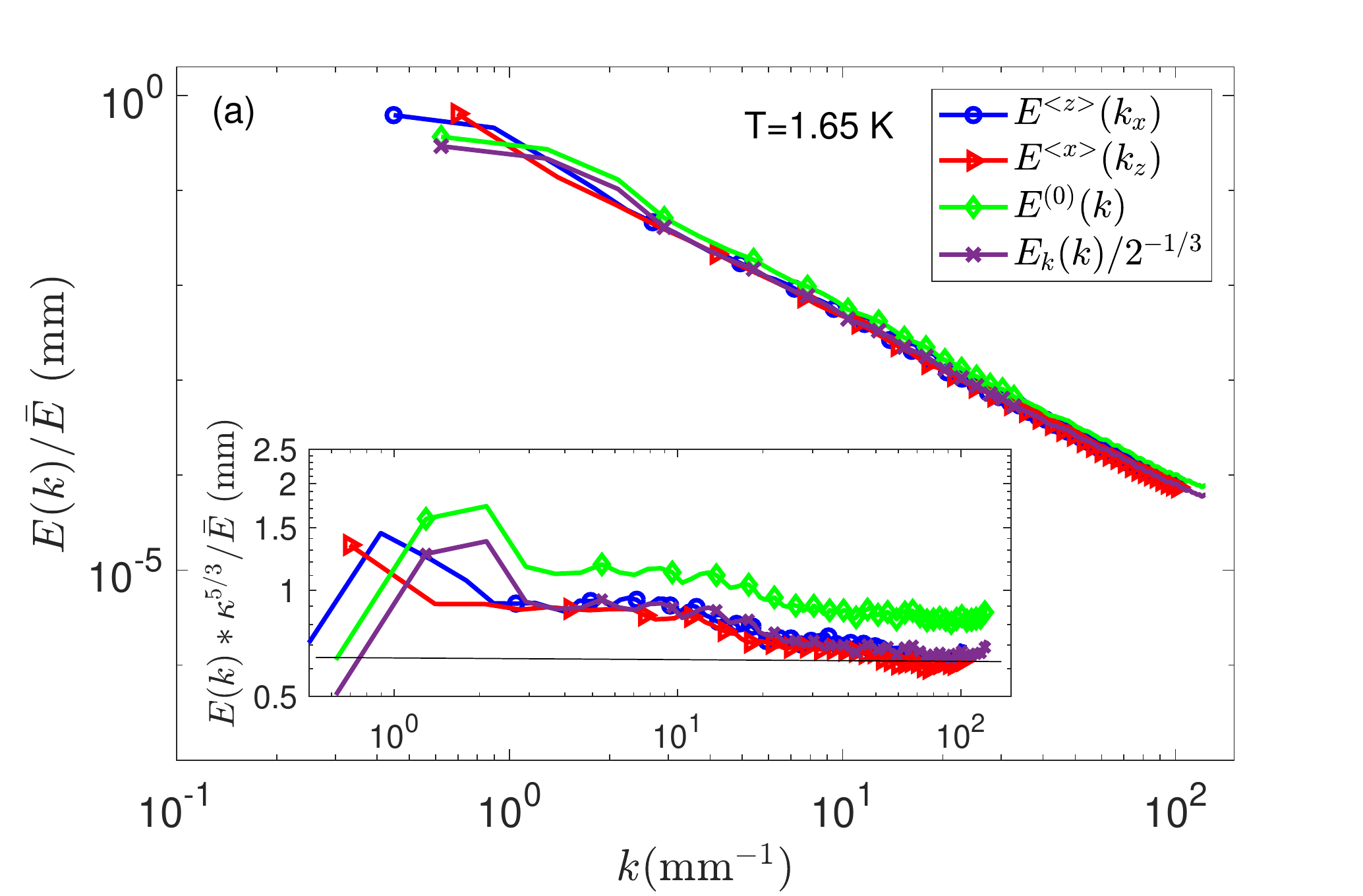} &
  		\includegraphics[width=.7\columnwidth]{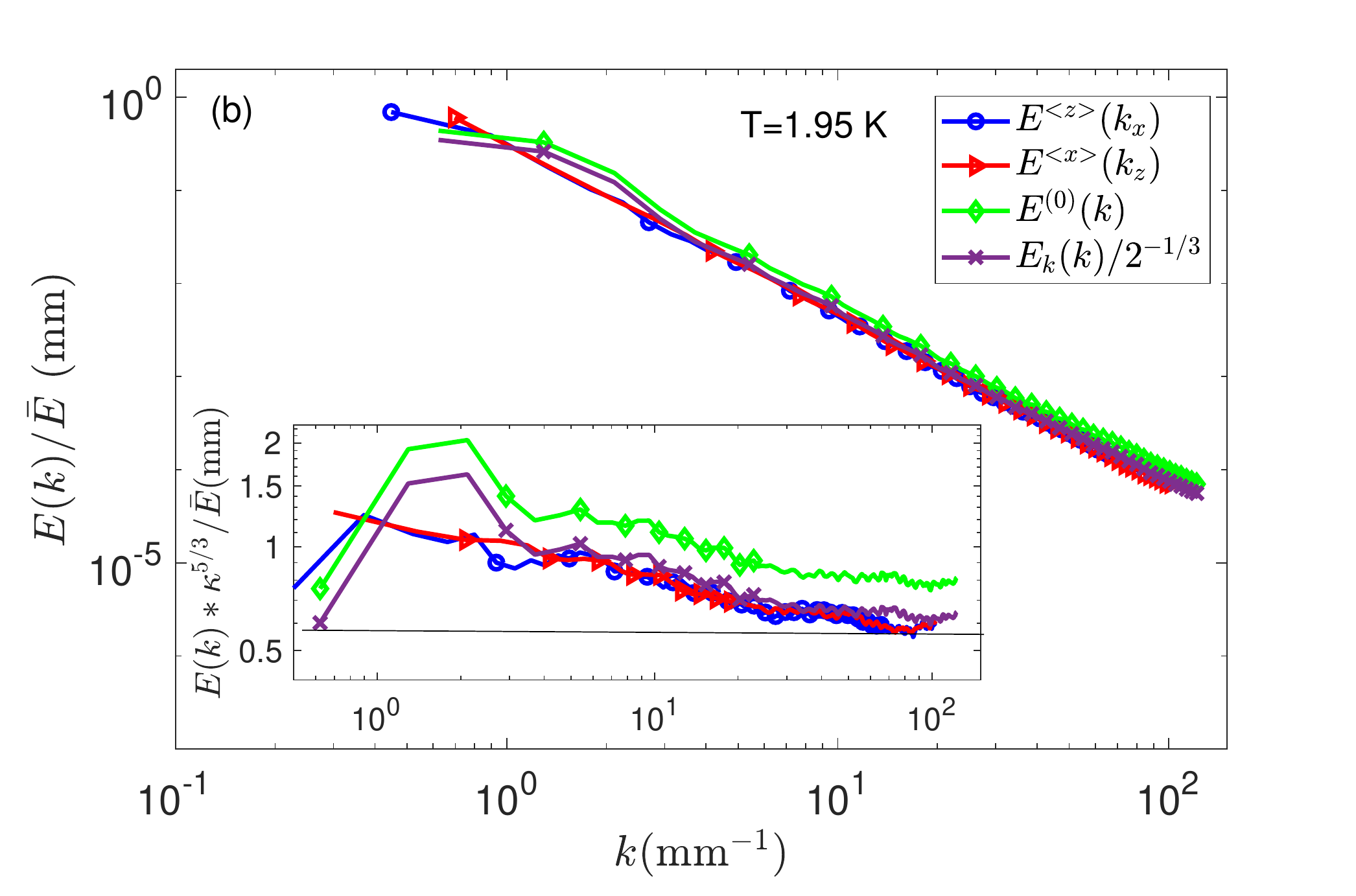} &
  		\includegraphics[width=.7\columnwidth]{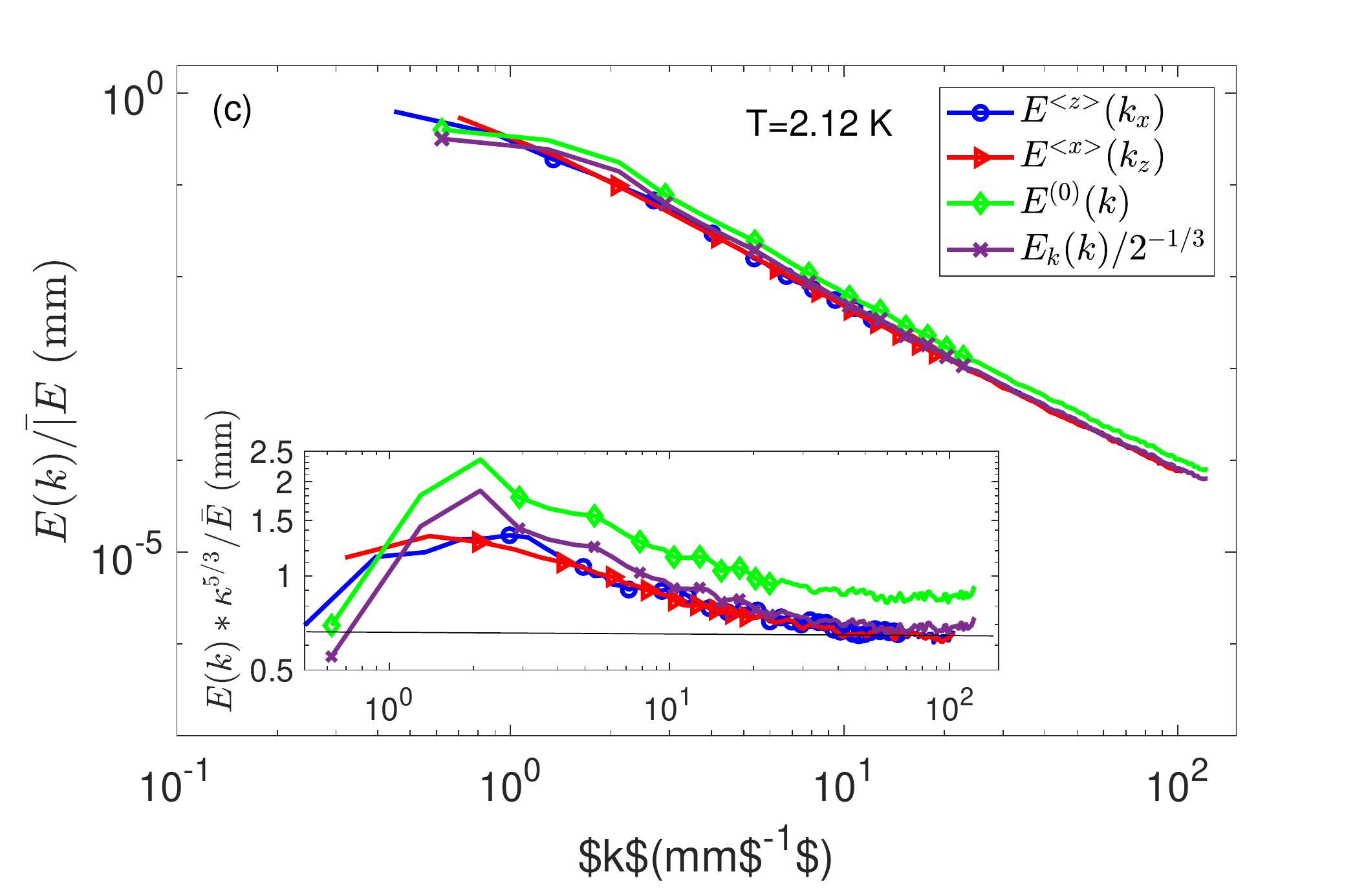} \\		
  		\includegraphics[width=.7\columnwidth]{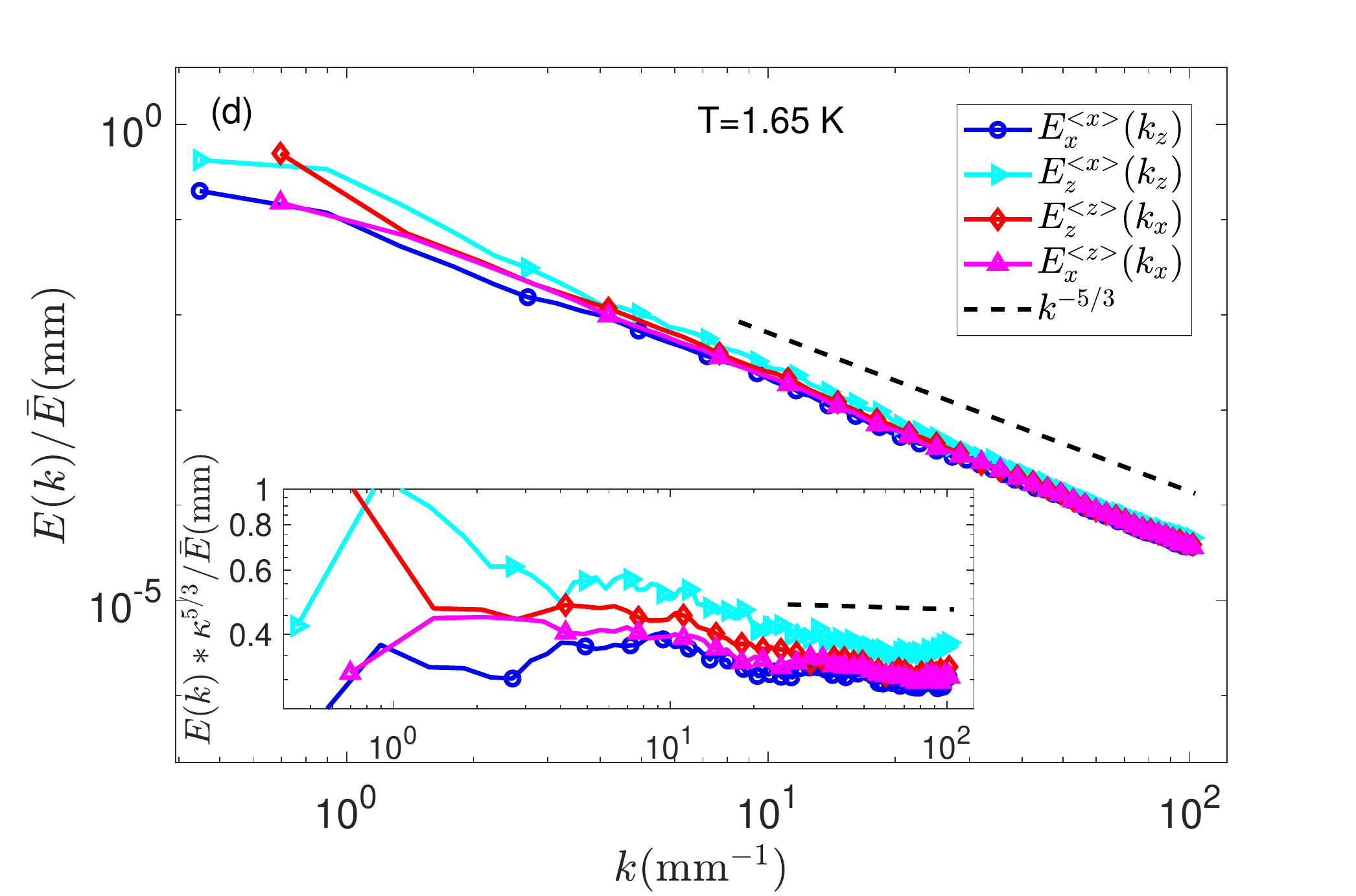} &
  		\includegraphics[width=.7\columnwidth]{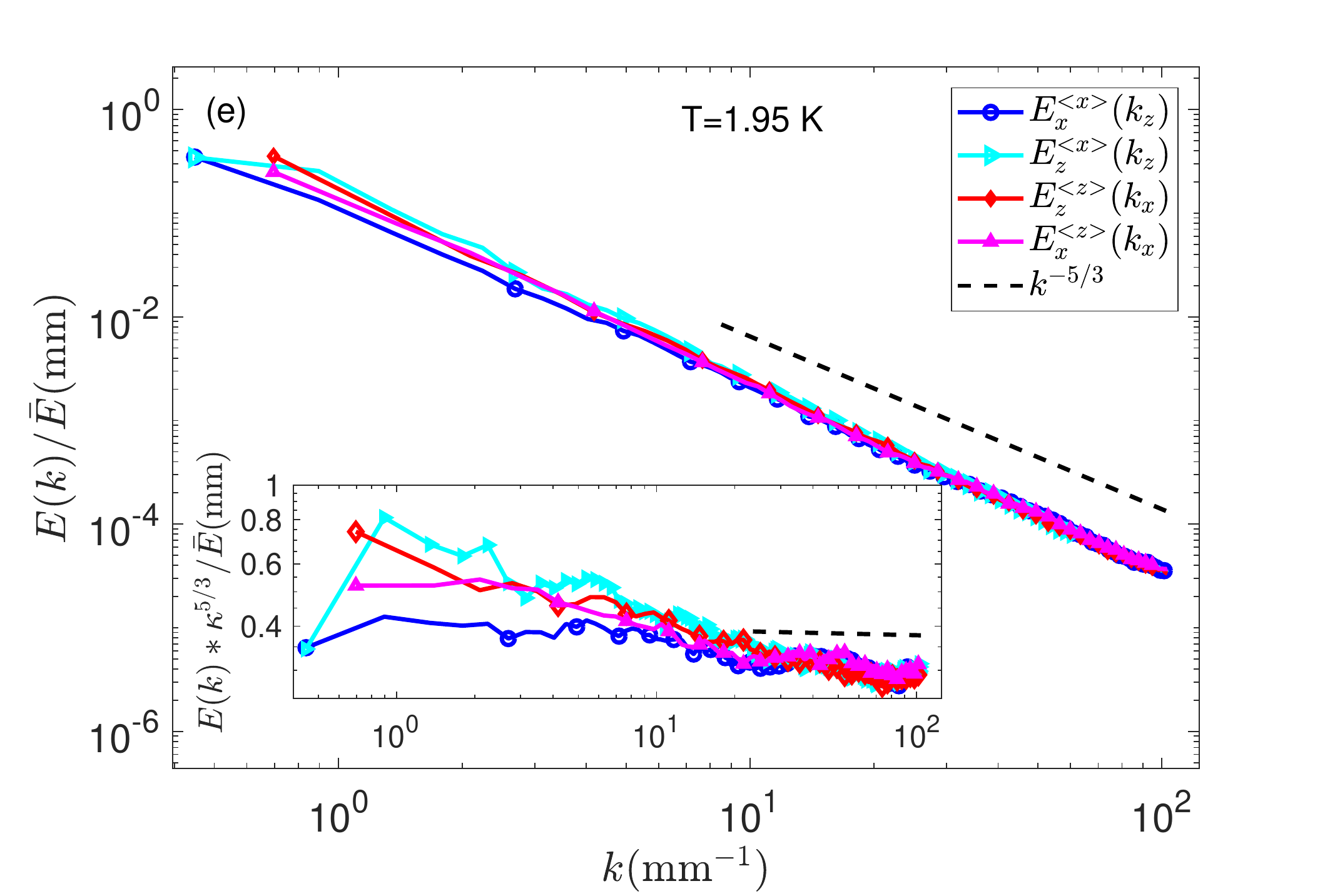} &
  		\includegraphics[width=.7\columnwidth]{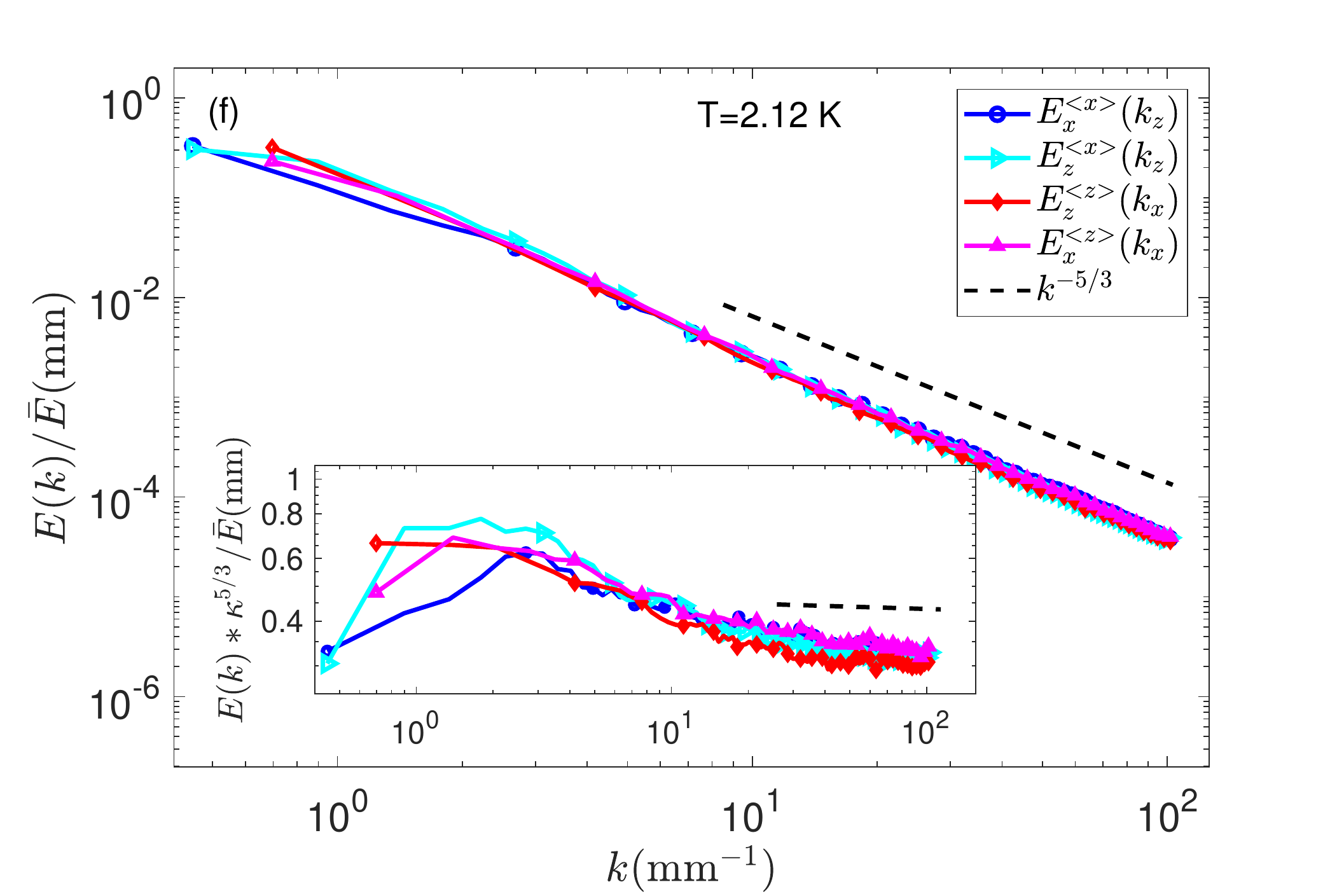}  \\
  	\end{tabular}
  	\caption{\label{f:lin}  (a), (b) and (c):   Comparison of     two linear energy spectra $E^{\langle x \rangle} (k_z) $ and  $E^{\langle z \rangle} (k_x)$ with angular-averaged  spectra   $E^{(0)}(k) $   for  $T=1.65\,$K,  $T=1.95\,$K and  $T=2.12\,$K. All spectra are normalized by total energy density per unit mass $\bar E$ for the given temperature.
  		 The horizontal gray lines mark (almost) temperature-independent asymptotic level of all normalized and compensated spectra $A\sb{as}\approx 0.65/\Delta k_x$. (d), (e) and (f): the individual components of the linear energy spectra. The spectra in the insets are compensated by K41 scaling $\kappa^{5/3}$ with   $\kappa\equiv  k/ \Delta k_x$, where $ \Delta k_x\approx 0.45\,$mm$^{-1}$. The K41 scaling is indicated by black dashed lines and serves to guide the eye only. 
  	}
  	
  \end{figure*}

\begin{figure*}[!ht]
	\begin{tabular}{ccc}	
		\includegraphics[width=1\columnwidth]{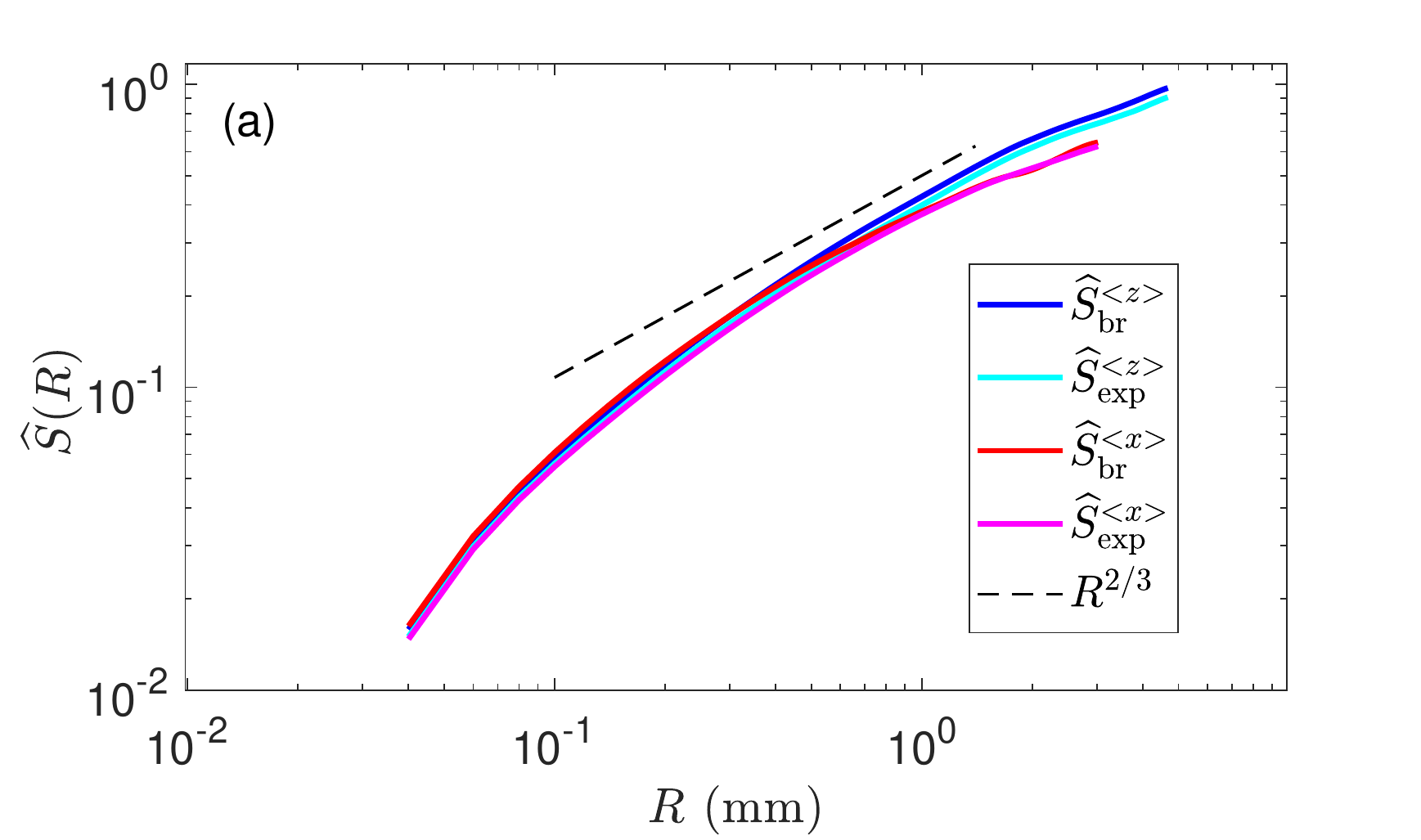} 
		\includegraphics[width=1\columnwidth]{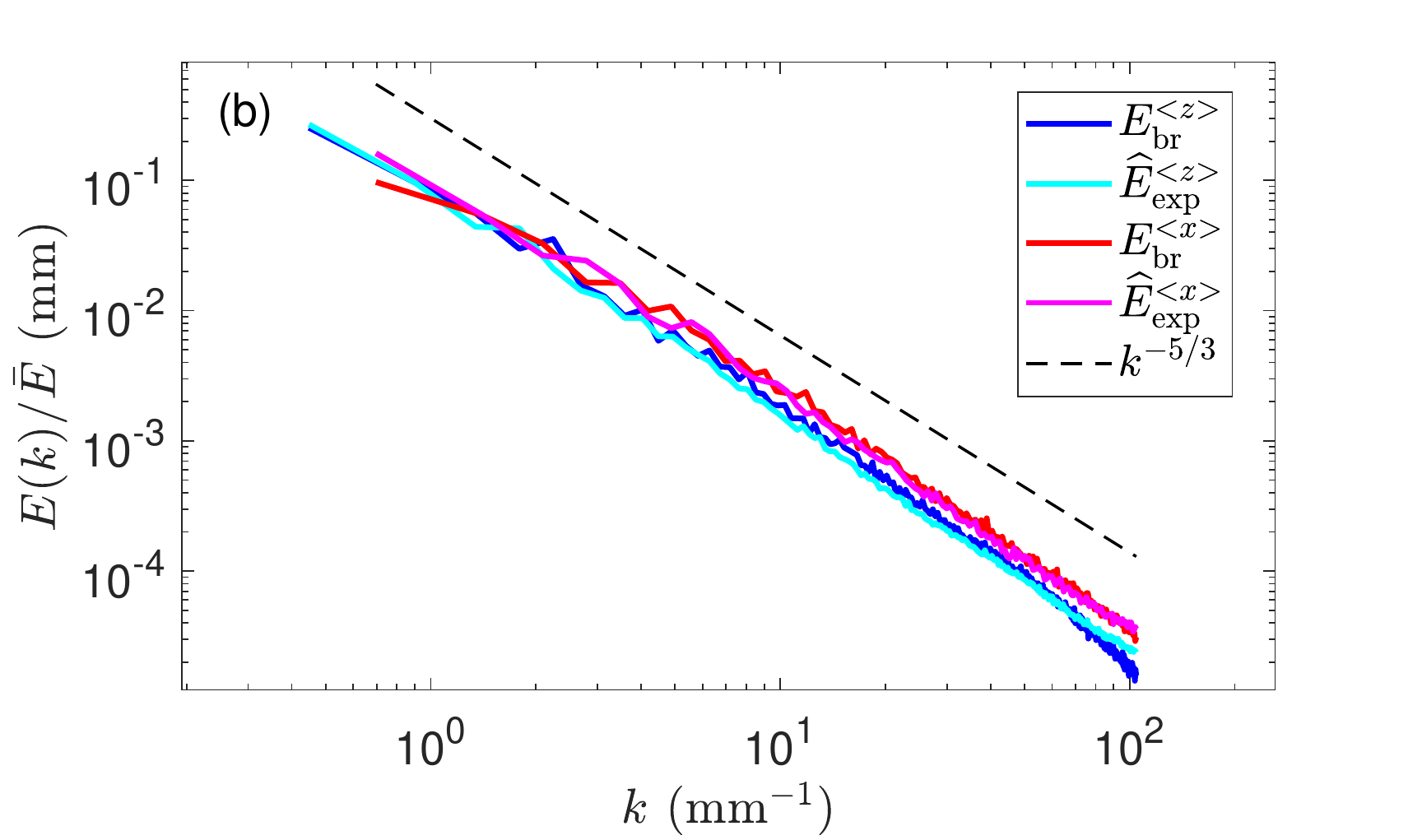} 	
	\end{tabular}
\caption{\label{f:bridges}
Experimental Eulerian  bridges for $T=1.95$\,K.
	 (a) The measured  $\widehat S^{\langle z \rangle}\sb{exp}$ and  $\widehat S^{\langle x \rangle}\sb{exp}$ compared with the  $\widehat S^{\langle z \rangle}\sb{br}$  and $\widehat S^{\langle x \rangle}\sb{br}$,  reconstructed from the measured energy  spectra using the bridge\,\Eq{4b}. (b) The measured  energy spectra $\widehat E\sb{exp}$ compared with $E\sb{br}$  reconstructed from the measured $\widehat C\sb{exp}$.  The black dashed lines, indicating the K41 scaling, serve to guide the eye only.
}

\end{figure*}

 In \Figs{f:lin}(a), (b) and (c)  we compare  angular-averaged energy spectra $E^{(0)}(k)$ with two linear spectra $E^{{\langle x \rangle}}(k_z)$ and $E^{\langle z \rangle}(k_x)$. All spectra are normalized by the energy  density $\overline E $ and  compensated by $\kappa^{5/3}$ with $\kappa \equiv k/\Delta k_x $.   As expected, all spectra in the inertial interval  (for $k>20\,$mm$^{-1}$ in our case)  have K41 scaling $\propto k^{5/3}$. We  see that two linear spectra almost coincide: $E^{\langle x \rangle}(k )\approx E^{\langle z \rangle}(k )$ confirming again the isotropy of the spectra. Note that  the angular-averaged  spectra $E^{(0)}(k)$, shown by green line, settle at larger values. The reason is that the K41 scaling is proportional to the energy fluxes: $E^{\langle x \rangle}(k_z )\propto \varepsilon _z^{2/3}$,  $E^{\langle z \rangle}(k_x )\propto \varepsilon _x^{2/3}$ and $E^{(0)}(k)\propto \varepsilon_k^{2/3}$. Here $\varepsilon _x\approx \varepsilon _z$ are  the energy fluxes  in the $\B {\hat x}$ and  $\B {\hat z}$-directions,
 while $\varepsilon _k= \sqrt {\varepsilon _x^2 + \varepsilon _z^2}\approx \sqrt 2 \varepsilon _x$. Therefore, we expect that $E^{(0)}(k)\approx 2^{1/3} E^{\langle z \rangle}(k )$. Indeed, $E^{(0)}(k)/ 2^{1/3}$ shown by cyan lines in
 \Figs{f:lin}(a)-(c), practically coincide with the plots  of  $E^{\langle x \rangle}(k_z)$ and $E^{\langle z \rangle}(k_x)$ for all temperatures.

Therefore, even in fully isotropic turbulence, 1D spectra obtained by different methods -- e.g. by one probe measurements with Taylor hypothesis of frozen turbulence, by axial averaging of two-dimensional spectra obtained via PIV or PTV method and by full averaging of three-dimensional spectra of turbulence --  all have different pre-factors, which need to be  accounted to compare experimental data  with theoretical or numerical results.

 Lastly, in \Figs{f:lin}(d), (e) and (f) we compare  linear spectra $E_\alpha^{\langle x \rangle}(k  )  $ and  $E^{\langle z \rangle}_\alpha(k)$ of the streamwise and wall-normal components  ($\alpha=x,\ z)$.  We see that for $T=1.95\,$K and $T=2.12\,$K all spectra practically coincide.  Only for $T=1.65\,$K the spectrum of $E^{\langle x \rangle}(k_z)$ is slightly more intense than other contributions. Moreover, at $T=1.65$ and $T=1.95$ K the energy components  measured in the streamwise direction differ  more at large scales  than those measured in the wall-normal direction, probably due to stronger anisotropy of the stirring force.

We thus conclude that Eulerian energy spectra of developed superfluid turbulence behind grid  are nearly isotropic   with respect of direction of the wave vector $\B k$ and   with respect of the  $x$- and $z$- vector projections of the velocity field in the available  part of the inertial interval  from $k\simeq 25\,$mm$^{-1}$ to $k\simeq 100\,$mm$^{-1}$. The finite spatial  resolution of the Eulerian approach does not allow us to determine  the  upper edge  $k\sb{max}$ of K41 scaling.  In \Sec{ss:Lagr} we will try solving this issue using Lagrangian analysis.

\subsubsection{\label{sss:br}  Experimental  Eulerian-Eulerian bridges} 
In \Sec{sss:bridges},  we formulated the Eulerian-Eulerian bridges\,\eqref{4b} and analyzed them in \Fig{f:1} using model energy spectra\,\eqref{mod} with  large inertial interval.   Here we demonstrate how bridges\,\eqref{4b} actually work for real  experimental data with a modest inertial interval.

 First of all, for each realization of the velocity field we calculate the structure and correlation functions using the same $\bar {\B  u}(\B R_{n,m})$ that was used to calculate the spectra, taking displacements $R_x$ along the $x$-direction and $R_z$ along the $z$-direction and averaging the resulting structure and correlation functions along the other direction. Next, we normalize them similar to \Eqs{norm}  and ensemble-average to get four objects: $\widehat S^{\langle z \rangle}\sb{exp}(R_x)$, $\widehat S^{\langle x \rangle}\sb{exp}(R_z)$, $\widehat C^{\langle z \rangle}\sb{exp}(R_x)$, $\widehat C^{\langle x \rangle}\sb{exp}(R_z)$. Finally, we use the normalized directly measured ``experimental" linear spectra \Eq{elin}
	\begin{equation}\label{Eexphat}
\widehat E\sb{exp}^{\langle z \rangle}(k_x)\=\frac{E\sb{exp}^{\langle z \rangle}(k_x)} {\bar E}\,, \quad \widehat E\sb{exp}^{\langle x \rangle}(k_z)\= \frac { E\sb{exp}^{\langle x \rangle}(k_z)} {\bar E}
	\end{equation}
  to reconstruct the structure functions $\widehat S^{\langle z \rangle}\sb{br}(R_x)$ and $\widehat S^{\langle x \rangle}\sb{br}(R_z)$ according to the bridge \Eq{4C}. Similarly,  we use  $\widehat C^{\langle x \rangle}\sb{exp}(R_z)$, $\widehat C^{\langle z \rangle}\sb{exp}(R_x)$ to reconstruct the linear spectra $E\sb{br}^{\langle x \rangle}(k_z)$, $E\sb{br}^{\langle z \rangle}(k_x)$, according to the bridge \Eq{4D}. In the latter calculations, to reduce the noise we use the data of the correlation functions between $R=0$ and the first minimum $R\sb{min}$ of $\widehat C\sb{exp}$  (not necessarily equal to zero), supplemented them with the same data, mirror-reflected around $R\sb{min}$ and used Fast Fourier Transform to calculate the spectra. The standard $2/3$ anti-aliasing rule was applied.  % The results are shown in \Fig{f:bridges}.
   Note that to compare the objects that depend on $x$- and $z$-directions, we accounted  for the respective increments in the calculation of the integrals and for the normalization factors, as in \Fig{f:lin}. We also limited the presented data by the $R$ and $k$ ranges available for the spectra, which are smaller than those accessible  by the structure and correlation functions. 

In \Fig{f:bridges}(a)   we compare the  ``experimental"	  $\widehat S ^{\langle x \rangle}\sb{exp}  $ and $\widehat S ^{\langle z \rangle}\sb{exp}  $ (cyan and magenta lines, respectively)  with the corresponding  $\widehat S ^{\langle x \rangle}\sb{br}  $ and $\widehat S ^{\langle z \rangle}\sb{br} $ (blue and red lines, respectively), 
reconstructed from the spectra \Eq{Eexphat}.  These spectra are shown in  \Fig{f:bridges}(b) by magenta and cyan lines and compared   with their  counterparts  $E\sb{br}^{\langle x \rangle} $ and $E\sb{br}^{\langle z \rangle} $ (red and blue lines, respectively). 
Note that our spatial resolution is about an order of magnitude larger than the Kolmogorov microscale (see Table \ref{table:KeyP}) and therefore the structure functions in \Fig{f:bridges}(a) do not reach the dissipative scales. As we showed in our previous work\cite{Bao-2018}, the structure functions with a finite (and relatively short) inertial interval demonstrate a gradual transition to an asymptotic scaling $R^2$ over a wide interval of scales. 

The overall agreement between the measured  and the bridge-reconstructed objects, demonstrated in  \Fig{f:bridges}, is very encouraging. This allows one to use bridge equations \,\eqref{4b}   as an efficient tool for the analysis of experimental and numerical results in studies of  hydrodynamic  turbulence at large as well as at modest Reynolds numbers.
 
In particular, we see that the 
 structure functions in \Fig{f:bridges}(a) at small scales do not depend on the orientation but  at large scale they differ. Both the measured $\widehat S\sb{exp}$ and the reconstructed  $\widehat S\sb{br}$ show a very narrow range of scales at which the scaling may be considered close to $R^{2/3}$. Over most of the available range, the scaling of the structure functions gradually changes from $R^{2/3}$ to $R^2$.  On the other hand, the reconstructed energy spectra $E\sb{br}$ exhibit a clear scaling close to $k^{-5/3}$ behavior over most of the wavenumbers range, similar to the $E\sb{exp}$ spectra, see \Fig{f:bridges}(b). 

 Therefore, as we mentioned in \Sec{sss:bridges}, the bridge relations between the velocity structure and correlation functions on one hand, and the energy spectra, on the other hand, do not require large inertial interval. The  only requirement is the   homogeneity of the velocity fields. 

 \subsection{\label{ss:Lagr} Lagrangian  statistics}
 \subsubsection{ \label{ss:LagrS}Lagrangian  2$\sp{nd}$-order structure functions}
  \begin{figure*}
  	\begin{tabular}{ccc}
		
 % 		(a)   & (b)   & (c)  \\
  		\includegraphics[width=.7\columnwidth]{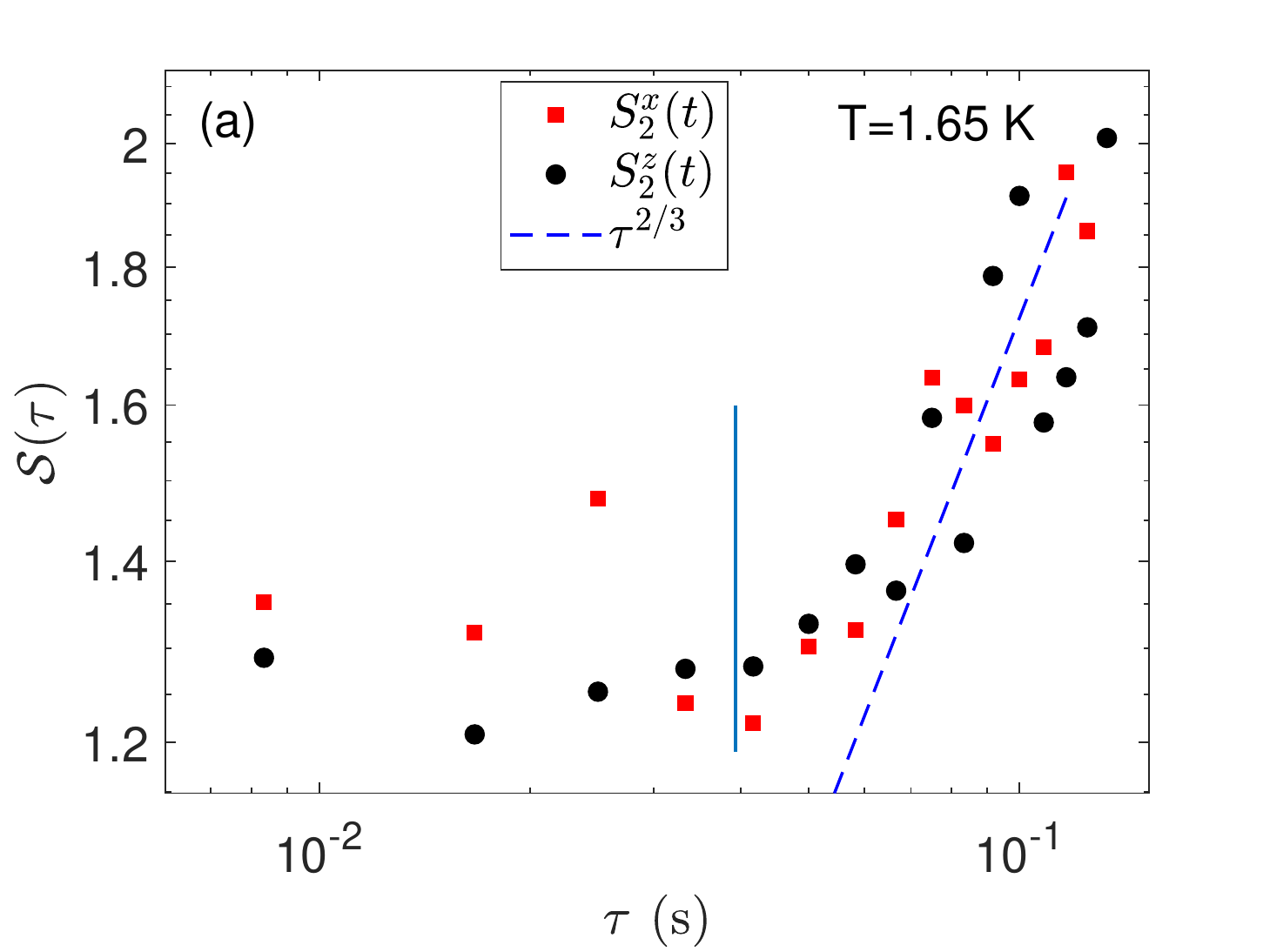}	&
  		\includegraphics[width=.7\columnwidth]{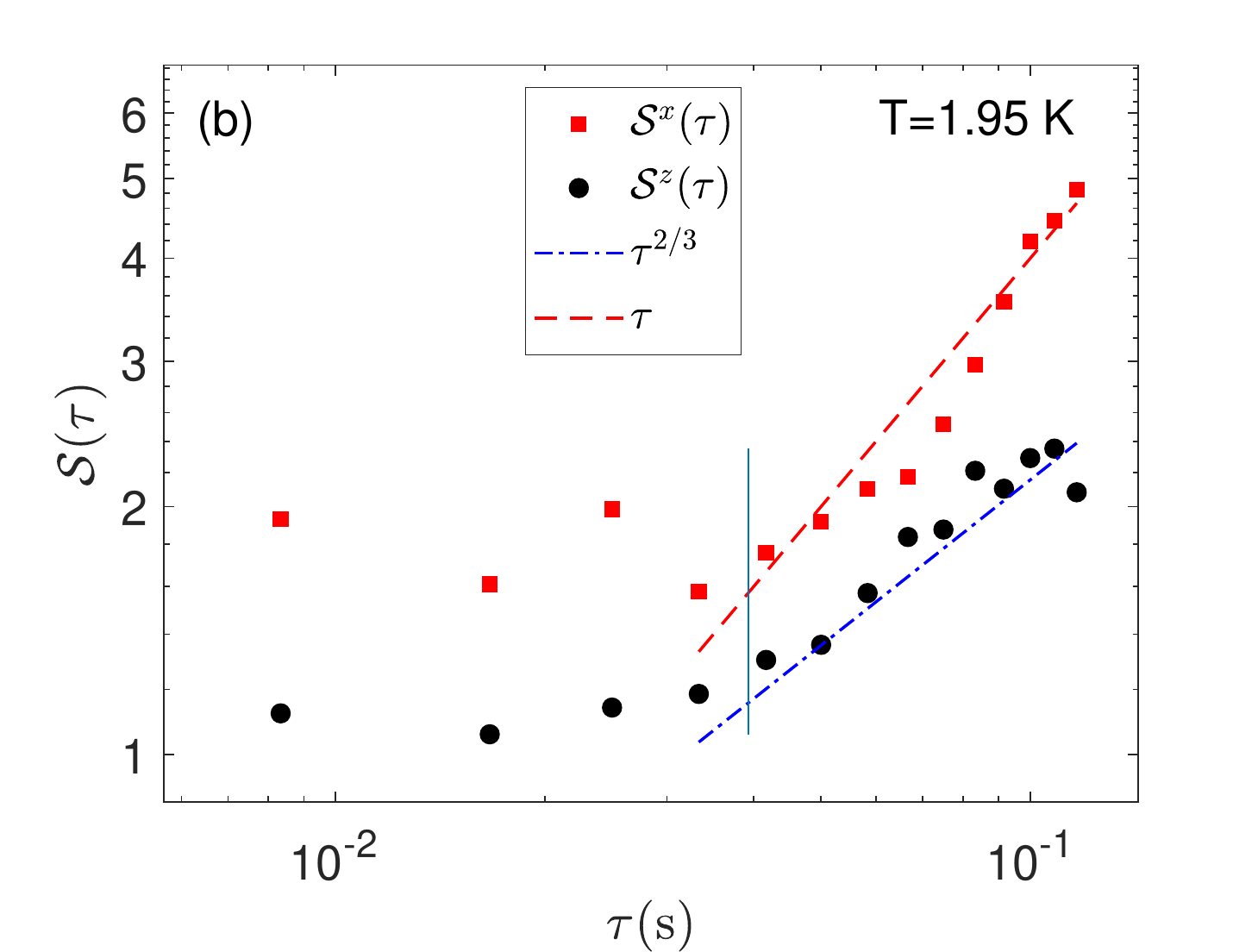}	&
  		\includegraphics[width=.7\columnwidth]{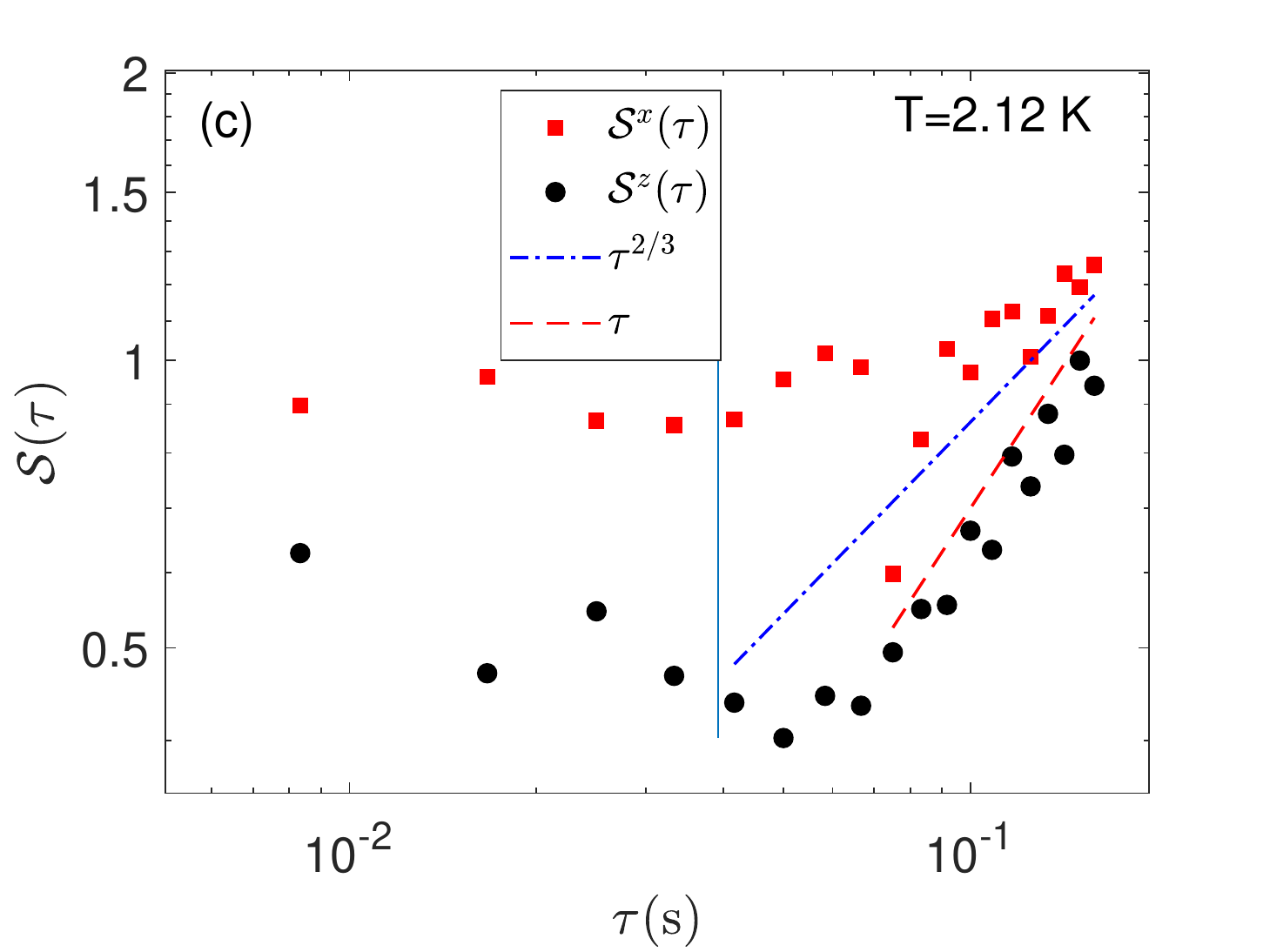} \\   && \\
  	%	(d)  & (e)   & (f)   \\
  		\includegraphics[width=.69\columnwidth]{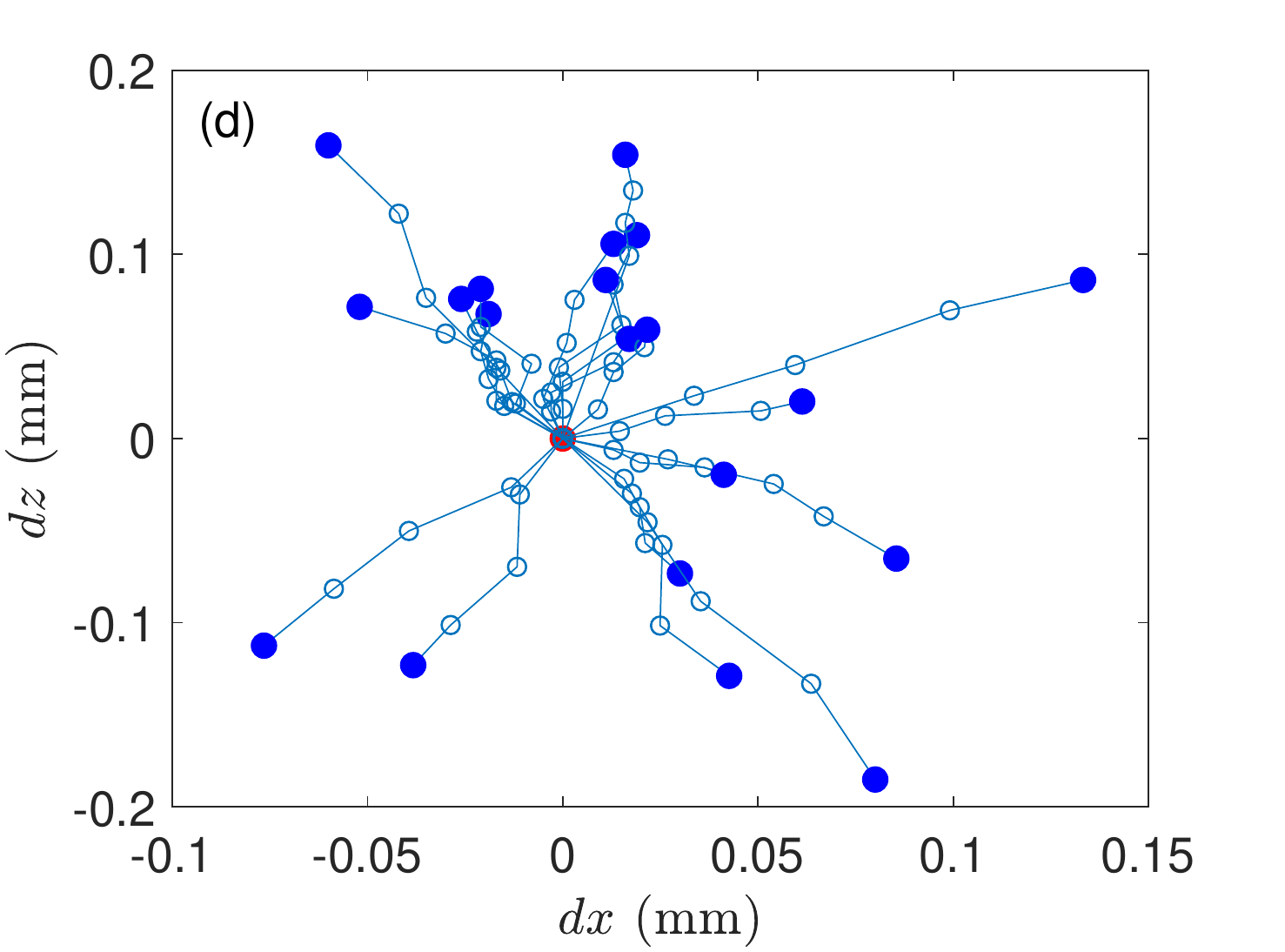}&
  		\includegraphics[width=.69\columnwidth]{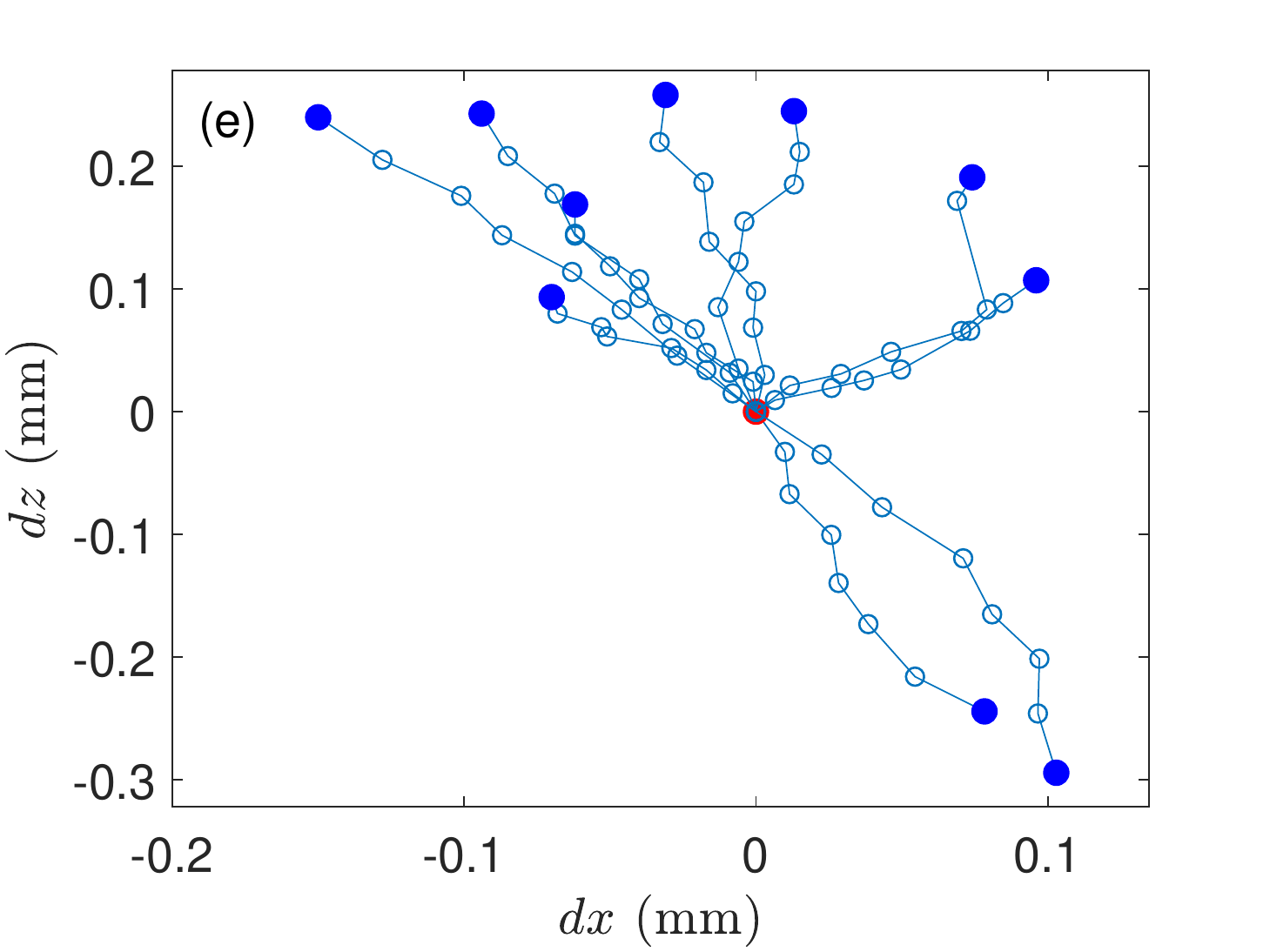}&
  		\includegraphics[width=.69\columnwidth]{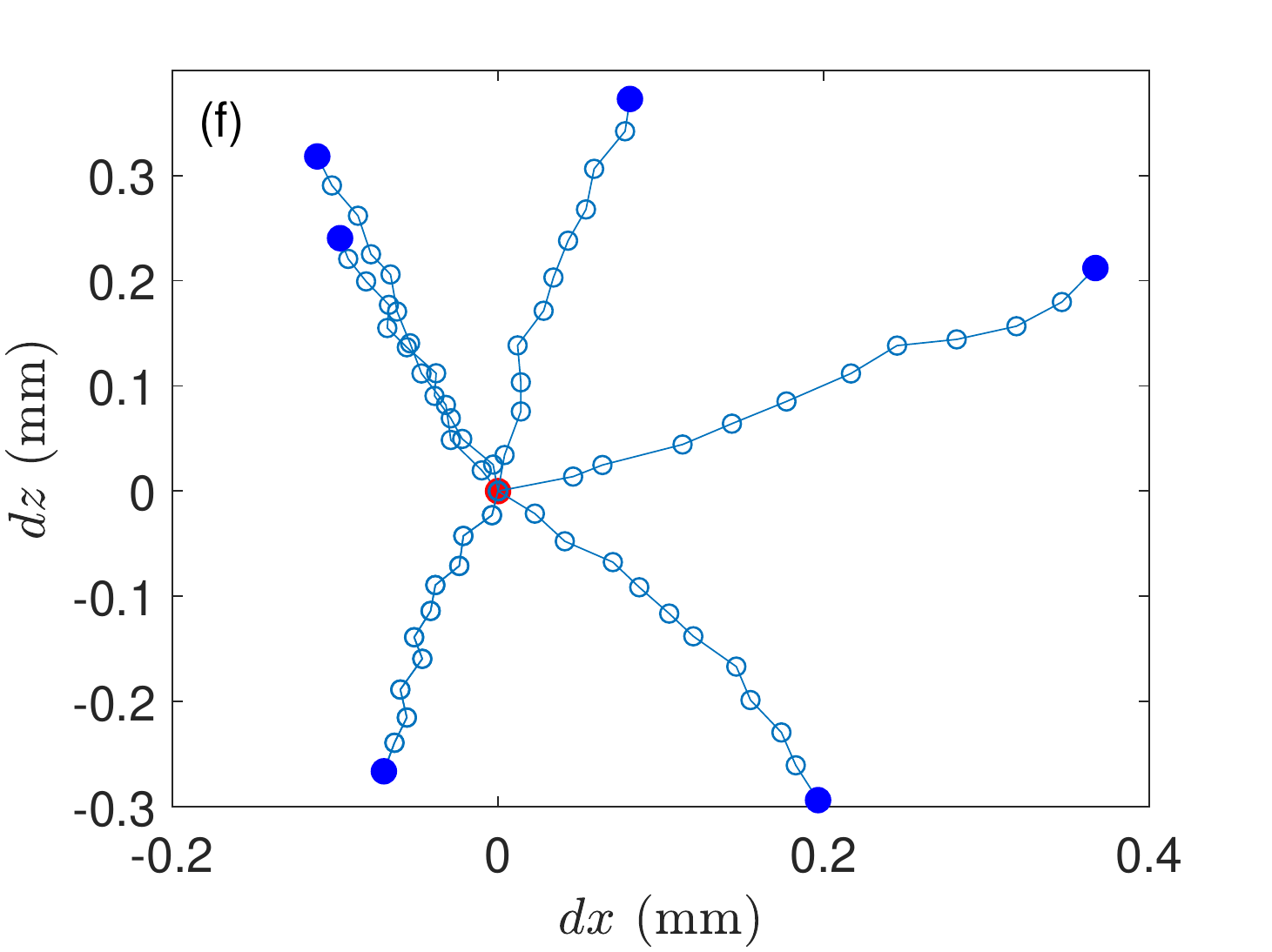}\\

  	\end{tabular}
  	\caption{\label{f:st} (a)-(c) Second order Lagrangian structure functions $\C S^{\alpha}(\tau)$ for different temperatures,  calculated using all trajectories with $N\geq 5$. Vertical lines denote $t\sb{cr}\simeq 0.04\,$s.
  	 (d)-f) 	Typical trajectories of particles  with 5, 8 and 12 points, respectively. The trajectories start at  the red point and end at the blue points. }

  \end{figure*}

\begin{figure*}
	\begin{tabular}{ccc}		
	%	(a) &(b) &(c) \\ 	
		 \includegraphics[width=.7\columnwidth]{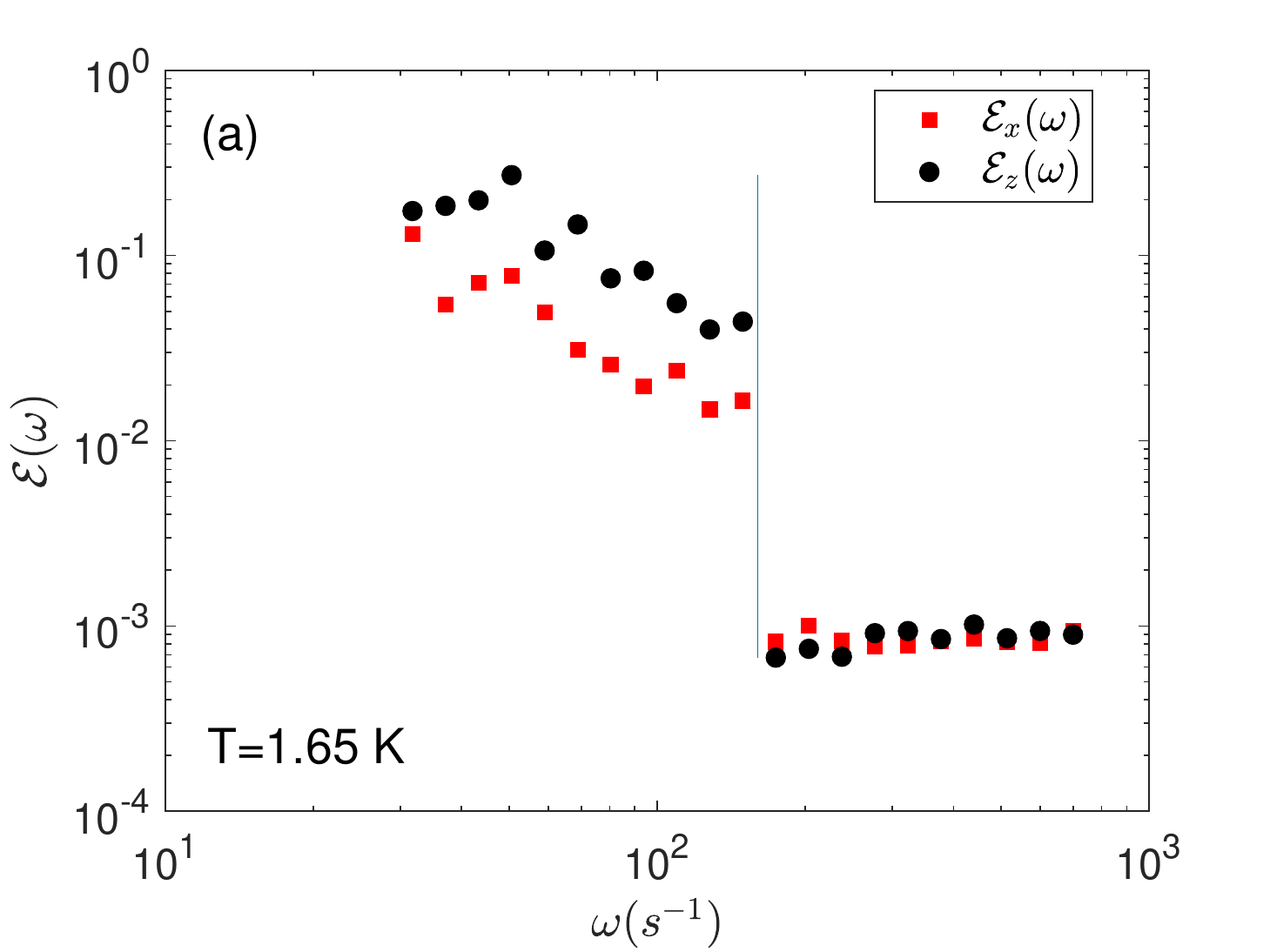}	&
		 \includegraphics[width=.7\columnwidth]{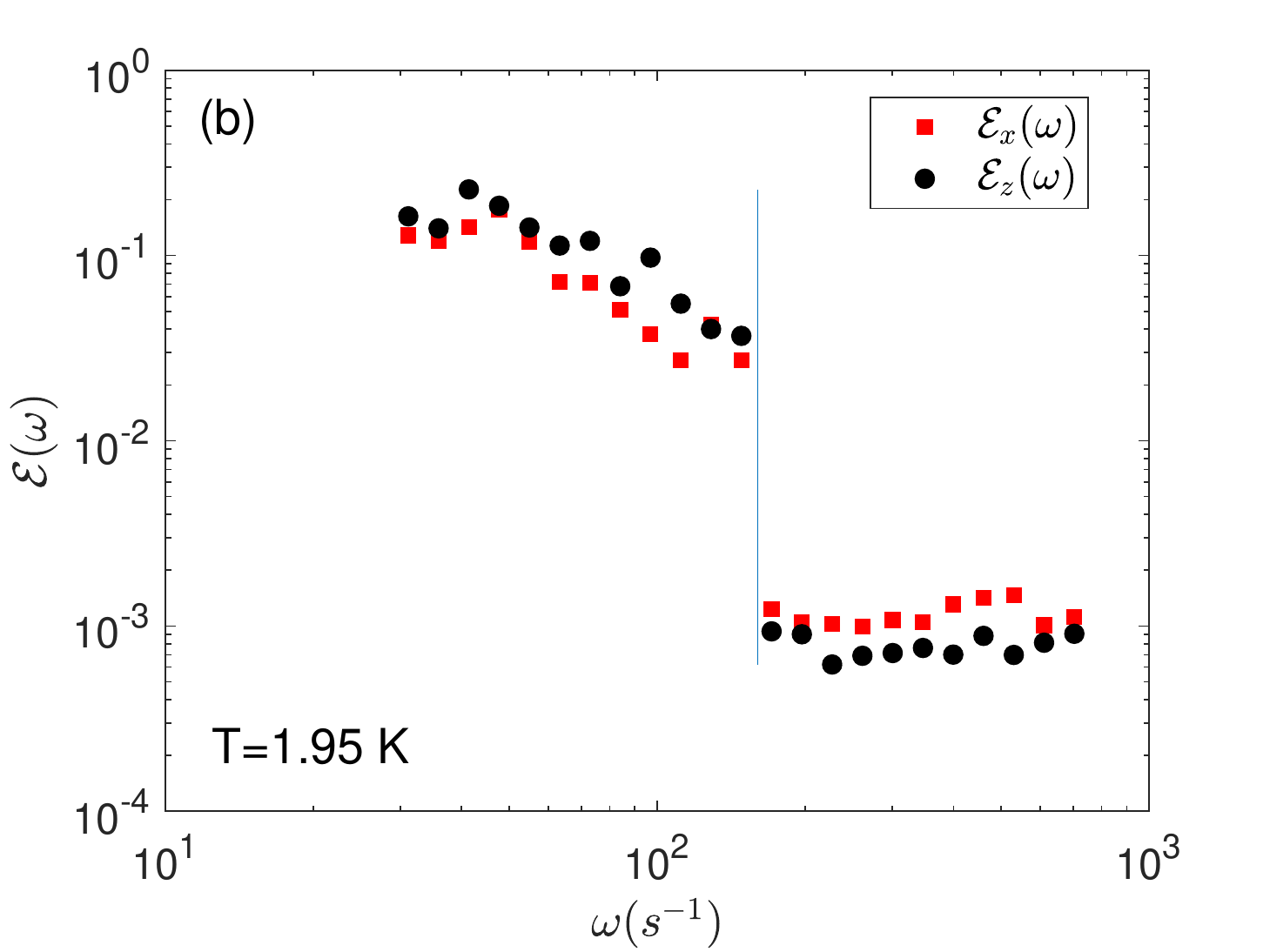}	&
		 \includegraphics[width=.7\columnwidth]{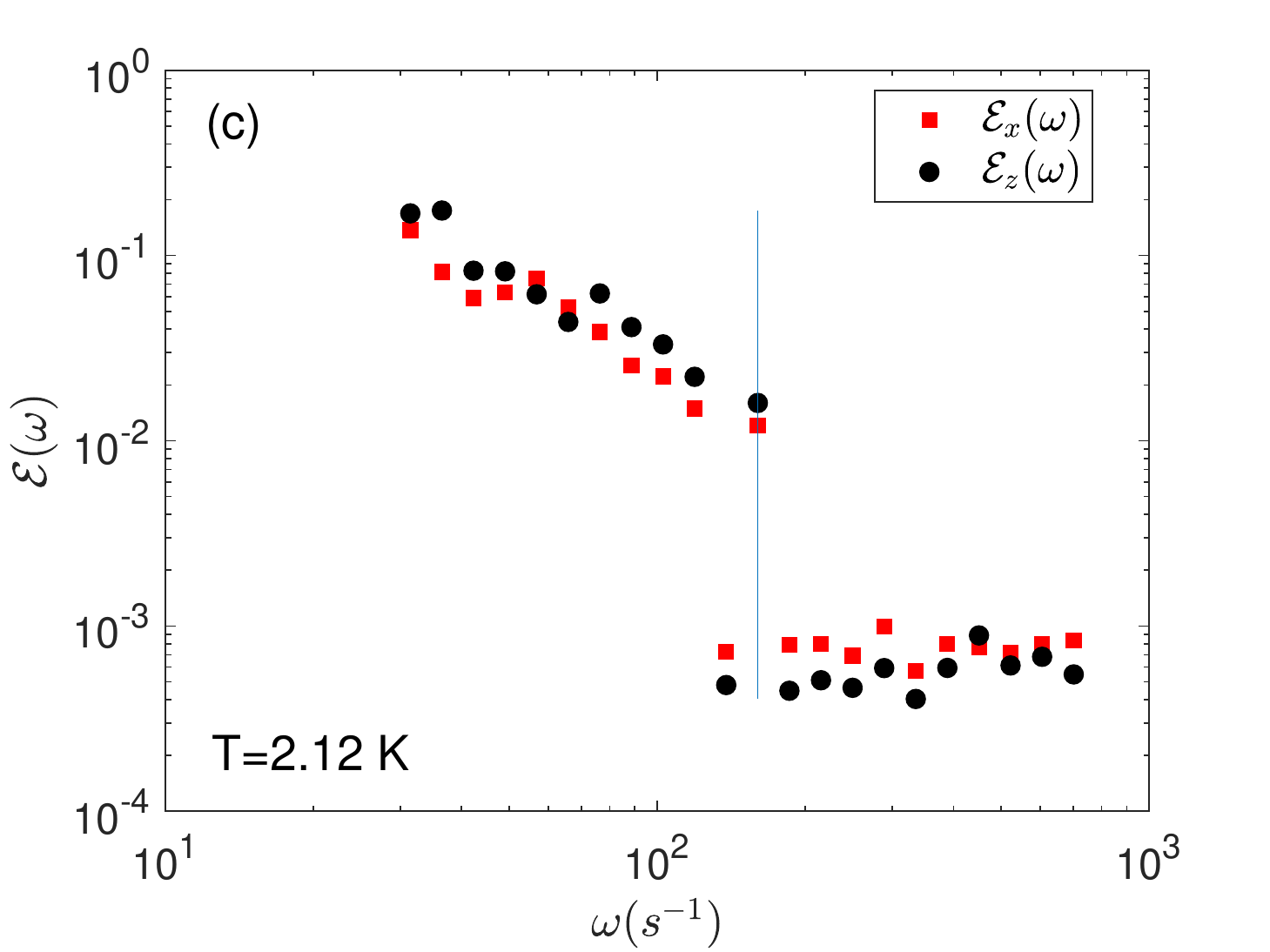} \\
	%	 	(d) &(e) &(f) \\ 	
		 \includegraphics[width=.69\columnwidth]{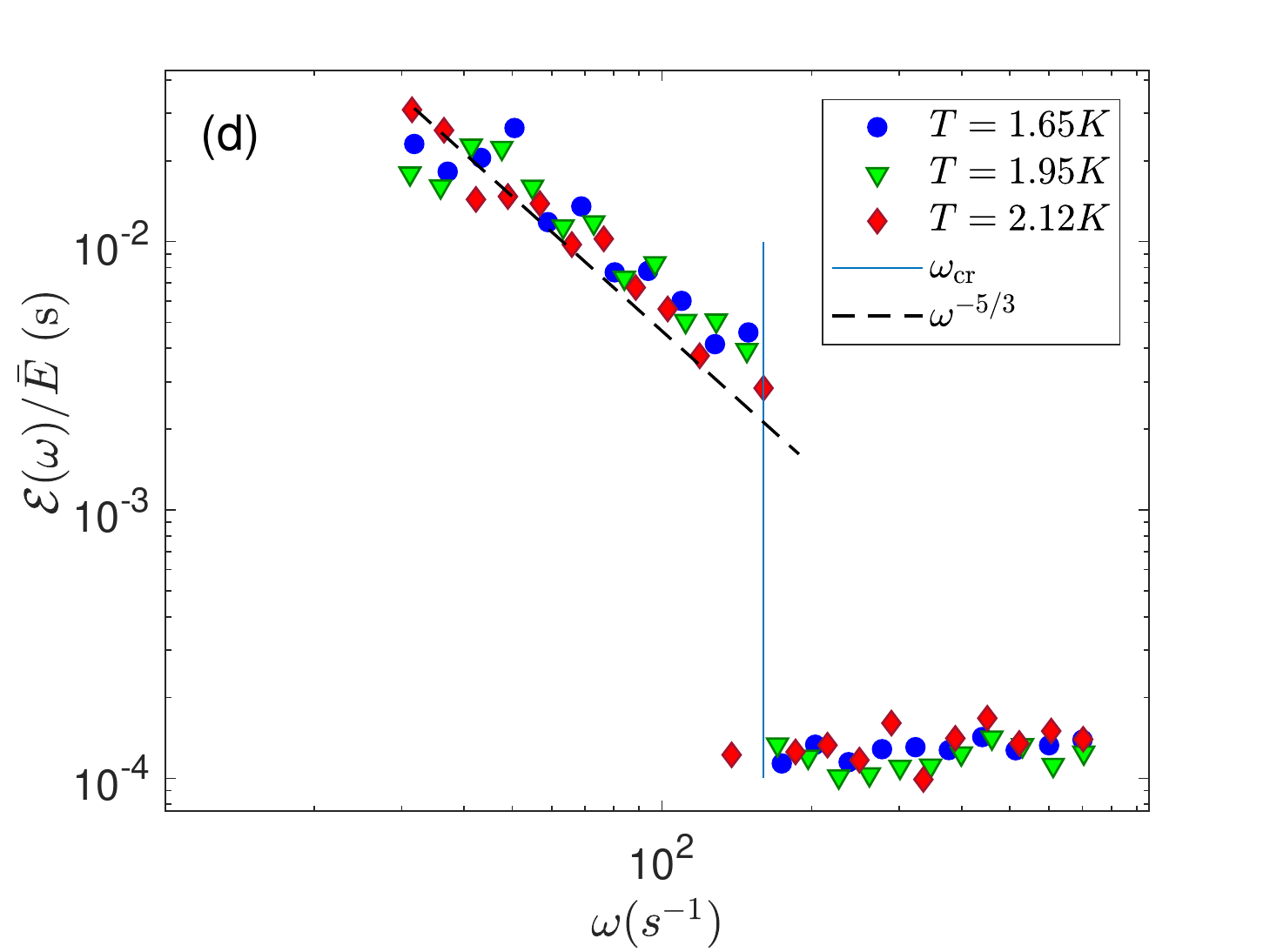}&
		 \includegraphics[width=.69\columnwidth]{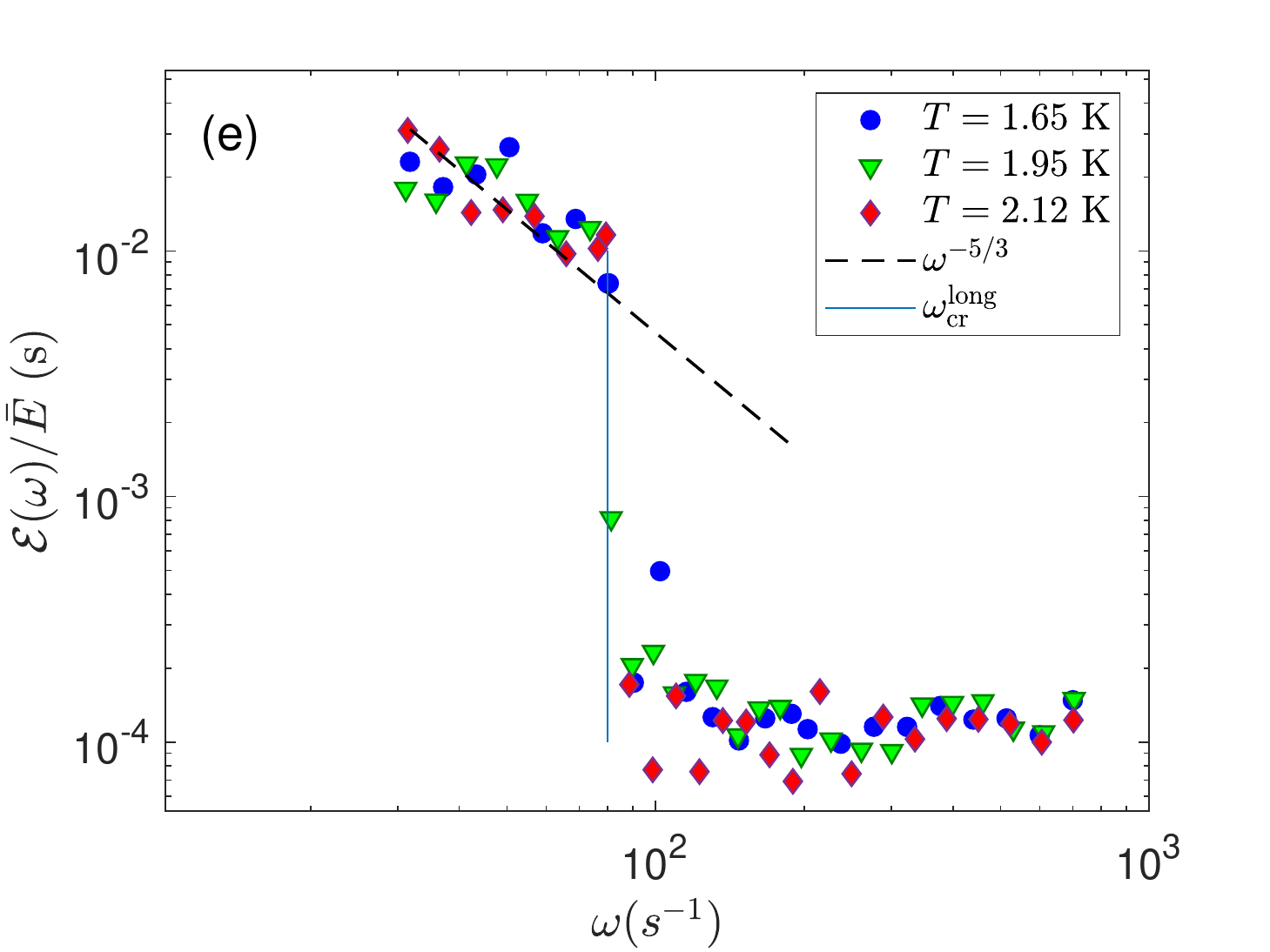}&
		 \includegraphics[width=.69\columnwidth]{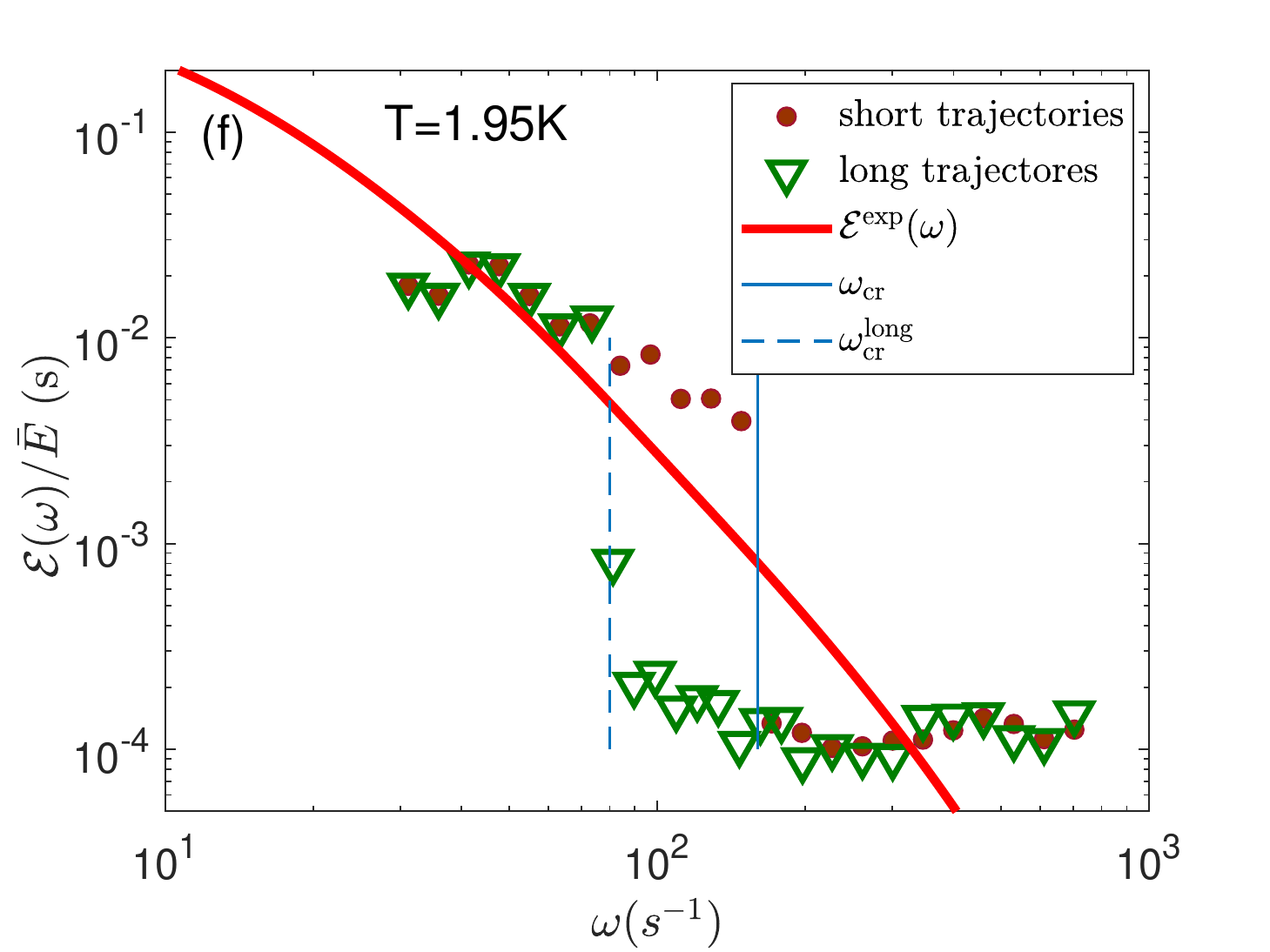}\\
	\end{tabular}
	\caption{\label{f:ew}(a)-(c) Lagrangian energy spectra $\C E_{\alpha}(\omega)$ at different temperatures, calculated using all trajectories with $N>5$.  $\C E_{x}(\omega)$ is marked by red squares and  $\C E_{z}(\omega)$ by the black circles. The  vertical line denotes the frequency corresponding to the cross-over time difference $\tau\sb{cr}\simeq 0.04$ s$^{-1}$.
	(d)-(e)  Normalized Lagrangian energy spectra $\C E(\omega)/\overline E$ for three temperatures. In (d) the spectra were calculated using only short ($N \in[5-9] $) trajectories, while in  (e) the spectra were calculated using only long ($N > 10 $) trajectories. (f)  Comparison of the normalized Lagrangian spectra $\C E(\omega)$= for $T=1.95\,$K calculated using short (brown dots) and long (green triangles) trajectories.  The thin vertical lines denote $\omega\sb{cr}=160$ s$^{-1}$,  $\omega^{\rm long}\sb{cr}=80$ s$^{-1}$. The red line denote the Lagrangian spectrum reconstructed from the Eulerian spectrum $E(k)$  using  the bridge \Eq{bridge}.
	% The red line corresponds to the original spectrum and the blue dashed line  corresponds to the spectrum extended by the K41 behavior $k^{-5/3}$ up to  $k=209$ mm$^{-1}$.
 }
\end{figure*}

Now we consider   second-order structure functions  of the Lagrangian velocity projections
	\begin{equation}\label{s2t}
	\C S ^\alpha(\tau)=\langle |u_j^\alpha(t+\tau)- u_j^\alpha(t)|^2\rangle_{t,j} \, ,\hfill \alpha=\{x,z\}\, ,
	\end{equation}
	averaged over all traces $j$ during all observation time $t$.  The structure functions are shown in \Fig{f:st} (a), (b) and (c).  We see that  their behavior is quite different from what is expected for  $\C S\sb{cl}(\tau)$ in classical hydrodynamic turbulence, shown in \Fig{f:smod}(c).    The only cases where the scaling of Lagrangian structure function in superfluid turbulence  coincides with that in classical K41 turbulence    $\C E(\tau)\propto \tau$ are    $\C S^z(\tau)$ for $T=2.12\,$K and   $\C S^x(\tau)$ for $T=1.95\,$K, see   red dashed lines in \Fig{f:st}(b) and (c).  In all other cases the $\C S (\tau)$ behavior, shown by blue dot-dashed  line in \Fig{f:st},  is closer to  the  scaling $\C S (\tau)\propto \tau^{2/3}$,    typical for classical K41 scaling of the Eulerian structure function, $S(r)\propto r^{2/3}$.
		
		To rationalize such an observation, we note that some particle trajectories, shown in \Fig{f:st} (d), (e) and (f), are quite close to straight lines with more or less equidistantly spaced points. Therefore, in these cases the velocities are measured similar to the Eulerian approach, i.e. the Eulerian scaling $S(r)\propto r^{2/3}$ transforms to the observed scaling $S(\tau)\propto \tau^{2/3}$. In some sense, this situation is similar to a one-point measurement of air turbulence in the presence of strong wind, where time-dependence of the turbulent velocity is transformed into $r$-dependence with the help of the Taylor hypothesis of frozen turbulence. The role of strong wind in our case is played by the energy-containing eddies with a random velocity which is much larger than the velocity of small-scale eddies in the inertial interval. The randomness of the sweeping velocity direction is clearly seen in  \Fig{f:st} (d), (e) and (f) as a random direction of the trajectories, that start at a red point and end at the blue points.

Another striking observation is that instead of  $\C S (\tau)\propto \tau^2$ behavior, originated from the smooth, differential velocity in the viscous range of turbulence in classical fluids, we observe a saturation of  $\C S (\tau)\approx$const. for small $\tau<\tau\sb{cr}\simeq 0.04\,$s. The independence of $\C S (\tau)$ from $\tau$ means that velocities  $\B v(t+\tau)$ and $\B v(t )$ are statistically independent. In this case \Eq{2a} gives
		\begin{equation}\label{sat}	
		\C S (\tau)= 2 \langle |\B v(t)|^2\rangle_t=\mbox{const for}\ \tau\lesssim \tau\sb{cr} \ .
		\end{equation}
		The simple physical picture of statistical  independence of $\B v(t+\tau)$ and $\B v(t )$  is based on the assumption that for $t\lesssim \tau\sb{cr}$ but still larger than the smallest time difference   $\Delta t=8\cdot 10^{-3}\,$s available in our experiments, the main contribution to the tracer velocity  consists of   sharp peaks uncorrelated during the time interval $\Delta t<  t\lesssim \tau\sb{cr}$.
	
	 A possible explanation\cite{WG1} is that for $r\lesssim \ell$  the normal and superfluid velocity components become practically decoupled. In this regime  the normal fluid turbulence is already damped by viscosity, while the superfluid  turbulence is supported by the random motion of the quantized vortex lines. Therefore, the motion of micron-sized particles in the range of scales $r\lesssim \ell$ is mainly controlled by the dynamics of the quantum vortex tangle,  including fast
	 events, such as vortex reconnections or particle trapping by the vortices.
	  This allows us to estimate the decorrelation time of their motion $\tau\sb{cr}$.
	
	The first step is to consider particles trapped on the vortex line. Their velocity can be estimated as
		the  root-mean-square velocity of the vortices $v_\ell$
		in the Local Induction Approximation\cite{Schwarz85,JGao-2018-PRB}
		\begin{subequations}\label{est}
			\begin{equation}\label{18}
			v_\ell \simeq   \frac{\kappa\Lambda}{4\pi \ell} \simeq   \frac{\kappa }{ \ell}\,, \quad \Lambda = \ln (\ell/a_0)\ .
			\end{equation}
			Here $\kappa \simeq 10^{-3}$cm$^2$/s is the quantum of circulation, $a_0\simeq 10^{-8}\,$cm is the vortex core radius. In \Eq{18} we have accounted for that at $\ell\simeq 0.05\,$mm the ratio $\Lambda/(4\, \pi)\approx 1.04\simeq 1$.   Then the decorrelation time $\tau\sb{cr}$ in this scenario can be estimated as the time  during which the configuration of the vortex tangle  changes significantly:
			\begin{equation}\label{19}
			\tau\sb{cr}\sim  \frac{\ell}{v_\ell}\simeq \frac{\ell^2}{\kappa}= \frac 1{\kappa \C L} \ .
			\end{equation}
	\end{subequations}
A possible role of the Magnus force in this scenario was considered in \Refs{Krst1,Krst2}. It   leads to the same estimate\,\eqref{19} for $\tau\sb{cr}$, which actually follows from the dimensional reasoning. 

To consider untrapped particles, we have to take into account that they directly interact with the superfluid component
		through the inertial and added mass forces\,\cite{SB}. Moreover,
		in the vicinity of the vortex core, the mutual friction induces, in the normal fluid,
		the vortex dipole whose typical lengthscale is expected\,\cite{Mastracci1-2019-PRF-1,SYui-2020-PRL} to be about 0.1\,mm, i.e. larger than $\ell$. It means that untrapped particles will be dragged by induced normal fluid component with  velocity about $\kappa/r$, where $r\lesssim \ell$ is the distance of the particle to the vortex core. For most of the untrapped particles $r\sim \ell$. In such a way,  we are coming to the same estimate\,\eqref{19} $\tau\sb{cr}\sim   \ell^2/\kappa$ for the untrapped particles as for the entrapped ones.
		With $\C L=4\cdot 10^4$\,cm$^{-2}$,
		\Eqs{18} gives the estimate  $v_\ell$ as 2\,mm/s and the decorrelation  time  $\tau\sb{cr}\simeq 0.025\,$s.  This time is  close to the critical time $\tau\sb {cr}\simeq 0.04$ s  below which $\C S(\tau)$ saturates, which thereby explains the saturation\,\eqref{sat} of  $\C S (\tau)$   for $\tau\lesssim \tau\sb{cr}$.

\subsubsection{\label{ss:LagrE} Lagrangian  energy spectra of turbulence}
In order to get information about the turbulence statistics at length scales below the cell resolution $\Delta$, we  compute the Lagrangian energy spectra $\C E(\omega)$ by analyzing the trajectories of individual particles. We first find the Lagrangian positions $\B X_n$ and velocities $\B {u}\Sb{P}(\tau_n)$ of a particle $P$ in the $(x,z)$-plane at consecutive moments of time $\tau_n=n \tau_0$. A Fourier transform in time of the velocity $\B {u}\Sb{P}(\tau_n)$, as described in \Sec{ss:Eul}, then gives the Fourier component $\B v\Sb{P}(\omega)$, which allows us to calculate the ensemble-averaged Lagrangian energy spectra $\C E(\omega)$:
\begin{equation}\label{22}
\C E(\omega)= \langle |\B v(\omega)|^2 \rangle\Sb{P}\, ,
\end{equation}
where the angled brackets now denote an average over an ensemble of particle trajectories $P$. Notice that  \Eq{22} is a discrete version of \Eq{5b} for $\C E(\omega)$.

The Lagrangian turbulent frequency power spectra $\C E(\omega)$ are shown in \Fig{f:ew}. The panels (a)-(c) show the energy spectra components
for $T=1.65\,$K, $T=1.95\,$K and $T=2.12\,$K, respectively. The most prominent feature of all spectra is a sharp fall by about two orders of magnitude at some $\omega\sb{cr}\simeq 160\,$s$^{-1}$  equal to $2\pi /\tau\sb{cr}$, where the critical time $\tau\sb{cr}\simeq 0.04\,$s 
 separates the  semi-classical regime (for $t>\tau\sb{cr}$) from the quantum regime, dominated by the velocity field induced by vortex lines (for   $t<\tau\sb{cr}$) in the behavior of $\C S^\alpha(t)$. Accordingly, the region $\omega>\omega\sb{cr}$ is expected to be mainly quantum, where we assume that the main  contribution to the tracers' velocity  consists of   sharp peaks uncorrelated in time. If so,  $\C E(\omega)$ should be $\omega$-independent for $\omega>\omega\sb{cr}$, as observed.

  Additional support for our scenario for the quantum-classical crossover frequency $\omega\sb{cr}$ is the estimate of turnover frequency of $\ell$-eddies of the intervortex separation scale $\ell$, $\omega_\ell\simeq 79.5-150.6\,$s$^{-1}$ for different $T$, which is quite close to the measured value $\omega\sb{cr}\simeq 160\,$ s$^{-1}$. It is commonly accepted that mechanically driven superfluid turbulence should behave almost classically for $\omega\ll \omega_\ell$ and  in a  quantum manner for  $\omega\gg \omega_\ell$.

 Notably, this transition does not happen at a single frequency. In \Fig{f:ew}(c) an overlap region is clearly seen, where both the classical and the quantum behavior coexists. It turns out that the velocity field calculated from shorter trajectories exhibit a longer classical frequency range, while for longer trajectories the transition is more gradual and starts at lower frequencies $\omega\sb{cr}\sp{long}$.  This behavior is temperature independent, see \Fig{f:ew}(d,e). Since the number of shorter trajectories is typically larger, the spectra that are ensemble-averaged over whole set of available trajectories are characterized by longer classical frequency range. We, therefore, expect that the actual transition occur in a range of frequencies  and is  not sharp. In \Fig{f:ew}(f)  we compare for $T=1.95\,$K  the spectra calculated using short and long trajectories (shown by symbols), and the Lagrangian spectra reconstructed from the Eulerian spectra  using the bridge \Eq{bridge} (thick line). The relation \eqref{bridge} is purely classical and does not describe the quantum  plateau in the spectra. Being normalized by the energy contained in the same frequency range $\omega\gtrsim 30\,$s$^{-1}$ as the experimental Lagrangian spectra, the reconstructed spectrum partially overlaps with the measured one in the "classical" range of frequencies, however with different scaling. The classical bridge has the expected $\omega^{-2}$ scaling, while the measured spectra scale as $\omega^{-5/3}$, as is shown in \Fig{f:ew}(d,e). This scaling matches the scaling of the structure functions, see \Fig{f:st}, and originates, as we suggested above, from  almost equidistant position of the tracers and consequent correspondence of $r$ and $\tau$ dependencies $E(r)\Leftrightarrow \C E(\tau)$ according to the Taylor hypothesis of frozen turbulence. Similar Lagrangian frequency power spectra with transition from $\omega^{-5/3}$ to  $\omega^{-2}$ scaling behavior  was observed (and explained), e.g,  in von Karman flows between counter-rotating  disks in \Refn{sweeping}. For more details about the interaction of particles with Kolmogorov turbulence  see \Refn{Krst2}.

  \section*{Conclusion}\label{s:sum} 
 
 In this paper we report a detailed analysis of the Eulerian and Lagrangian second-order statistics -- the velocity structure and correlation functions together with the energy spectra -- measured by the particle tracking velocimetry  in the superfluid $^4$He grid turbulence in a wide temperature range. 
 
We measured two-dimensional Eulerian spectra $F(k_x,k_z)$ in the $(\B x,\B z)$-plane, oriented along the streamwise $\B z$-direction.  Using SO(2) decomposition, we demonstrate that  the plane anisotropy of the studied grid turbulence is very small and can be peacefully neglected.  This allows us to further analyze only one-dimensional energy spectra. We use three of them: the angular averaged spectrum $E^{(0)}(k)$ and two ``linear" energy spectra $E^{\langle x \rangle }(k_z)$ and $E^{\langle z \rangle }(k_x)$, averaged over the corresponding direction. We show that with a simple and physically motivated renormalization of the energy fluxes, these three spectra practically coincide.  Independent of the way of averaging, the Eulerian energy spectra have extended inertial scaling range with close to $k^{-5/3}$ behavior. The available range of the  Eulerian spectra, however, does not allow to probe the transition to the viscous or quantum regime. This was achieved by analysis of the Lagrangian structure functions and spectra.   These demonstrate the sharp transition from a near-classical behavior to a time- and frequency independent plateau, respectively. The transition occurs at a range of time increments and frequencies that are consistent with the intervortex scale, defined by the measured vortex line density.  The appearance of such a plateau corresponds to the statistically independent velocities of the tracer particles associated with the velocity field dominated by the velocities of the quantum vortex lines. The scaling behavior in the Lagrangian  spectra in the quasi-classical regime $\omega<\omega\sb{cr}$ deviates from the expected K41 $\omega^{-2}$ behavior and is closer to $\omega^{-5/3}$. We suggest that the possible origin of this discrepancy is the shape of the particles trajectories.  Many of them are almost straight and particles positions are almost equidistant at the subsequent measurement.  Such a situation is well described by a Taylor hypothesis of frozen turbulence, in which  the time dependence of the measured velocity  effectively corresponds to a one-time $r$-dependence. Hence the scaling typical to the Eulerian spectra.
 However, this unexpected scaling behavior near the classical-quantum transition requires further investigation.
 
 The unique feature of PTV measurements, allowing to simultaneously extract from the same set of tracers' velocities the Eulerian and Lagrangian statistical information, allowed us to verify the set of bridge relations that connect various statistical objects, both within the same framework  (Eulerian-Eulerian and Lagrangian-Lagrangian) and connecting two ways of the statistical representation (Eulerian-Lagrangian).

 In particular,  we demonstrate using the experimental data that two-way bridge \Eqs{4b} between the Eulerian energy spectrum $E(k)$, the velocity structure $S(r)$ and the correlation functions $C(r)$ allows us to reconstruct with high accuracy any two of these objects using remaining one of them for any extend of the inertial interval including very modest one.  Similar equations\,\eqref{6}  connect the Lagrangian structure functions and the power spectra.  These bridges may be used as the efficient tool for the analysis of experimental and numerical data in studies of hydrodynamic turbulence at any Reynolds numbers  opening 
 multi-sided view on the statistics of turbulence. For example, for modest  Reynolds numbers, if $S(r)$ does not have a visible scaling regime, $E(k)$  may reveal its  existence. The correlation function $C(r)$ stresses large-scale properties of turbulence including possible  coherent structures, while $S(r)$ highlights the small and moderate scale behavior of turbulence.  
 
 We also demonstrate how  the  combination of Eulerian-Lagrangian bridge  \Eqs{6} and \eqref{bridge} allows one to reconstruct the Lagrangian second-order statistical objects -- the energy spectrum $\C E(\omega)$, the structure and the correlation function from the Eulerian spectrum $E(k)$ in classical turbulence.  The bridge  \Eq{bridge}  does not describe the transition to the quantum regime.  Unfortunately, this is one-way bridge: one cannot find $E(k)$ from $\C E(\omega)$ without additional model assumptions. 

 We hope that further improvements in PTV techniques  together with other possible methods of superfluid velocity control will allow one to  explore the  intervortex  range of scales in more detail and to get more information about the quantum behavior of superfluid turbulence.

\end{document}